\documentclass[reprint,amssymb,amsmath,superscriptaddress]{revtex4-2}

\usepackage[utf8]{inputenc}

\DeclareUnicodeCharacter{2212}{\textendash}

\usepackage[T1]{fontenc}
\usepackage{amsmath,amssymb}
\usepackage{amsfonts}
\usepackage{graphicx}
\usepackage{float}
\usepackage{wasysym}
\usepackage{url}
\usepackage{booktabs}
\usepackage{svg}
\usepackage{physics}
\usepackage{mathtools}
\usepackage{siunitx}
\usepackage{wrapfig}
\usepackage{lipsum}
\usepackage{enumitem}
\setlist[enumerate]{itemsep=0mm}

\begin{document}

\title{The stabilizing role of multiplicative noise in non-confining potentials}

\author{Ewan T. Phillips}
\email{tphillips@pks.mpg.de}
\affiliation{Max Planck Institute for the Physics of Complex Systems, Nöthnitzer Str. 38, 01187 Dresden, Germany}

\author{Benjamin Lindner}
\affiliation{Bernstein Center for Computational Neuroscience, Haus 2, Philippstr. 13, 10115 Berlin, Germany}
\affiliation{Department of Physics, Humboldt Universität zu Berlin, Newtonstr. 15, 12489 Berlin, Germany}

\author{Holger Kantz}
\affiliation{Max Planck Institute for the Physics of Complex Systems, Nöthnitzer Str. 38, 01187 Dresden, Germany}

\date{\today}

\begin{abstract}
We provide a simple framework for the study of parametric (multiplicative) noise, making use of scale parameters. We show that for a large class of stochastic differential equations increasing the multiplicative noise intensity surprisingly causes the mass of the stationary probability distribution to become increasingly concentrated around the minima of the multiplicative noise term, whilst under quite general conditions exhibiting a kind of intermittent burst like jumps between these minima. If the multiplicative noise term has one zero this causes on-off intermittency. Our framework relies on first term expansions, which become more accurate for larger noise intensities. In this work we show that the full width half maximum in addition to the maximum is appropriate for quantifying the stationary probability distribution (instead of the mean and variance, which are often undefined). We define a corresponding new kind of weak sense stationarity. We consider a double well potential as an example of application, demonstrating relevance to tipping points in noisy systems. 

\end{abstract}

\maketitle

\section{Introduction}
Multiplicative noise (or parametric noise) is ubiquitous in real world systems and (in contrast to additive noise) is known to play an important role in state transitions \cite{horsthemke1984noise, feudel1997multistability, pisarchik2014control, feudel2008complex, kraut2002multistability, kraut1999preference, huerta2008experimental, de2007noise}, including tipping points where hopping is induced from one attractor to another \cite{landa2000changes, forgoston2018primer, van1994noise, hodgkinson2021multiplicative}. Such events have relevance for example in synchronization \cite{kim1997noise, kim1997reentrant, lee1998phase, boccaletti2002synchronization}, human balancing tasks \cite{cabrera2002off}, financial crises \cite{krawiecki2002volatility, mandelbrot1997variation}, environmental tipping points \cite{imkeller2001stochastic, RN4, majda1999models, majda2009normal, sardeshmukh2009reconciling, berner2017stochastic, sura2005multiplicative, buizza1999stochastic}, neuroplasticity \cite{wang2022multiplicative, bauermann2019multiplicative}, epidemics, and stochastic optimization \cite{pavlyukevich2007levy}. 

Another important noise induced phenomenon is that of on-off intermittency, where noise induces an aperiodic switching between static, so called laminar behavior and chaotic bursts \cite{platt1993off, heagy1994characterization, ding1997stability, elaskar2023review}. This is due to the noise causing the bifurcation parameter to fluctuate around the bifurcation point \cite{ashwin1994bubbling}. Such behavior is often observed as a continuous route from regular behavior to chaotic motion \cite{elaskar2023review, hramov2006ring, scheffer2009early}. Such intermittency has been observed for example in noisy laser systems, which may exhibit sporadic high intensity pulses\cite{young1988statistical, zhu1993steady, chowdhury2022extreme}, Earth surface temperature at both weather and climate time scales \cite{schertzer1987physical, lovejoy2018spectra, ashwin2012tipping, sura2002noise} and synchronization of coupled systems of interacting dynamical units, where the bursting induces a phase slip \cite{syncphillips23, gilpin2021desynchronization}. 

Theoretical work on the topic of noise induced transitions has generally taken two different approaches. In the first approach the effect of noise on a system is quantified according to the moments (mean, variance etc.) of the probability distribution \cite{schenzle1979multiplicative, aumaitre2007noise, bourret1973brownian, lindenberg1981brownian, lucke1985response}. This approach, while useful in a small noise or additive noise limit, is however not able to describe noise induced bifurcations, since unbiased noise does not affect the mean. The effect of noise on a system is instead to broaden and/ or to skew the distribution, generally leading to heavy tails and in some cases even leading to divergence of the moments. In the second class the sharp transition in the presence of noise is interpreted as the bifurcation of a maximum of the stationary probability density of a distribution \cite{horsthemke1984noise}. It remained unexplained, however, why only the maximum of the distribution should be observed in experiments \cite{graham1982stabilization}. 

In this paper we explore a class of Langevin stochastic differential equations (SDE's), which we show under quite general conditions exhibit on-off intermittency. We build on the second approach suggesting the use of the maximum as an alternative to the mean and the Full Width Half Maximum (FWHM) as an alternative to the variance for systems with multiplicative noise. These two quantities collectively account for the body of the distribution excluding the heavy tail. We find counter intuitively that by skewing the tail of the distribution the noise induces a kind of stabilization of the minima of the noise term. 

It has been known for some time that unbiased parametric noise may stabilize an autonomous linear stochastic system of two or more dimensions in the sense of the moments, and that such noise is not sufficient to stabilize a one dimensional SDE \cite{arnold1983stabilization, arnold1979influence, bobrik1999stabilizing}. Here, however, we show that noise is sufficient to stabilize a one dimensional SDE in the sense of the maximum and the FWHM.

The structure of this text is as follows. We first explore a general class of SDE's. We then focus on the noise stabilizing case where the trajectory turns out to exhibit on-off intermittency. We show how these systems can be analyzed in terms of scale parameters and how from this the FWHM can be derived. We then show why the mean is insufficient to arrive at conclusions about the systems with a finite number of trajectories. In the following section we explore the tails by considering first passage times (distribution of the bursts). We finish by generalizing the results and as an example of application exploring the effect of multiplicative noise in a double well potential. 

\section{Basic model}
The general one dimensional Langevin SDE is
\begin{equation}
    \dot{x} = f(x) + \sqrt{2D}g(x)\xi(t), \label{SDE}
\end{equation}
where $\xi(t)$ represents Gaussian white noise with mean $\langle \xi(t) \rangle = 0$ and correlation $\langle \xi(t)\xi(t') \rangle = \delta(t-t')$. $f(x)$ and $g(x)$ represent the drift and diffusion terms respectively. We focus on the case where the drift $f(x)$ represents a non-confining potential, i.e. the trajectories drift toward infinity in the absence of noise. Due to the symmetry of the noise around zero and its lack of temporal correlations we notice that we may replace $g(x)$ with $|g(x)|$, and thus assume $g(x) \geq 0$. For what follows we furthermore assume that $g(x)$ has only one zero. The probability distribution of Eq. (\ref{SDE}) evolves according to the Fokker Planck equation given by 
\begin{equation}
    \frac{\partial}{\partial t} p(x, t) = - \frac{\partial}{\partial x}[f(x)p(x, t)] + \frac{\partial^2}{\partial x^2}[Dg(x)^2 p(x,t)].
\end{equation}
We first consider the case with natural boundaries. The stationary distribution is easily obtained from the condition $\partial_t p(x, t) = 0$ as  
\begin{equation}
    p_{s}(x) = \frac{N}{g^{2}(x)}\exp{2\int^{x}\frac{f(u)}{g^{2}(u)}du} \label{stat_dis_abs}
\end{equation}
The extrema $x_m$ of the stationary solution $p_{s}(x)$ of the Fokker Planck Equation (FPE) can be easily obtained (by setting the derivative to zero $p'_{s}(x) = 0$)  as \cite{horsthemke1984noise}
\begin{equation}
    f(x_m) - (2-\nu)D g(x_m)g'(x_m) = 0. \label{max}
\end{equation}
Here $\nu = 1$ refers to the Stratonovich interpretation and $\nu = 0$ to the Itô interpretation. From this equation we see immediately that as $D$ increases the term $Dg(x)g'(x)$ becomes dominant, and thus that the $x_m$ moves closer to the zero of $g(x)$ as $D$ increases. This suggests that for large $D$ it may be sufficient to take the first term of a Taylor expansion around the solution of $g(x_0) = 0$.

We expand $g(x)$ as a series and take the first non zero term of the expansion $g(x) \approx a_{m}x^m$ (where $m$ is the order of the lowest non zero term of the series expansion). The details of the higher order terms will often be unimportant, since we will be primarily interested in the case that the orbit spends most of the time in the vicinity of the minimum of $g(x)$. Considering for the moment that $f(x)$ has a fixed point and $g(x)$ has a minimum at $x = 0$ (generalizations are discussed in the appendix) suggests studying the general class of first order Langevin SDE's 
\begin{equation}
    \dot{x} = ax^n + \sqrt{2D}x^{m}\xi(t) \label{sde_monomial}
\end{equation}
where the drift $a \in \mathbb{R}_{>0}$ acts to create a non-confining potential. $n, m \in \mathbb{R}$ although we will generally restrict ourselves to (real numbers) $n, m \leq 1$ in order to guarantee that the solution does not explode (see appendix). We assume here natural boundary conditions (this choice will also prove to be unimportant in the vicinity of $x = 0$). It can be obtained easily from Eq. (\ref{max}) that the maximum of the stationary distribution changes with $D$ according to 
\begin{equation}
    x_m = \left(\frac{(2-\nu)m D}{a}\right)^\frac{1}{n-2m+1}. \label{max scale}
\end{equation}
We now obtain by straightforward calculation of the second derivative that the calculated extremum of Eq.~(\ref{sde_monomial}) is a maximum ($p''_{s}(x_{max}) < 0$) in both Itô and Stratonovich interpretation if and only if the condition $n - 2m + 1 < 0$ holds. We note that these equations also holds exactly even if there is additional additive noise in Eq. (\ref{sde_monomial}). The stationary distribution of Eq. (\ref{sde_monomial}) is given by
\begin{align}
    p_{s}(x;\beta) &= Nx^{\nu-2m}\exp{\frac{a}{D}\frac{x^{n-2m+1}}{n-2m+1}}.  \label{stat dist}
\end{align}
This distribution is not always normalizable. The normalization constant is calculated in Itô interpretation as
\begin{align}
    N^{-1} &= \begin{cases}
      \text{const.} & \text{if $n - 2m < -1$}\\
      \infty & \text{if $n - 2m \geq -1$}.
    \end{cases}  \label{self_sim_norm}
\end{align}
This result also holds in Stratonovich interpretation provided $m > 1$ (See appendix for details). We now restrict ourselves to the study of Eq. (\ref{sde_monomial}) in Itô interpretation $\nu = 0$, although we remark that qualitatively both interpretations lead to the same basic results. We have so far seen that for $n - 2m + 1 < 0$ the stationary distribution has a single maximum and is normalizable. We now state furthermore in this regime that this distribution has a scale parameter $\beta$, i.e.
\begin{equation}
    p_s(x;\beta) = \frac{p_s(x/\beta; 1)}{\beta}.  \label{scaling dist 2}
\end{equation}
with scaling parameter given by 
\begin{equation}
    \beta = \left(D/a \right)^{\frac{1}{n-2m+1}}. \label{beta}
\end{equation}

\begin{figure}
	\begin{center}
		\hspace*{-1cm}\includegraphics[width=9.3cm]{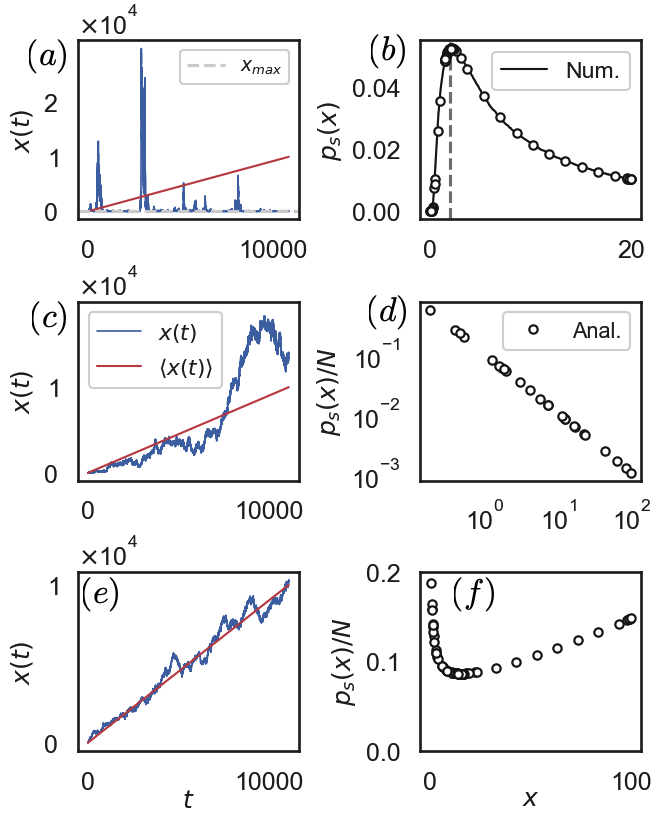}\caption{Time series and (unnormalized) stationary probability distribution for Eq.~(\ref{sde_monomial}) where $n = 0, a = 1$ and $m = 0.8, D = 0.4$ (a)-(b), $m = 0.5, D = 1$ (c)-(d), $m = 0.35, D = 5$ (e)-(f). For (d) and (f) $p_{s}(x)/N$ plotted (where $N$ is the normalization constant), since $p_{s}(x) = 0$.
		}\label{m03}
	\end{center}
\end{figure}

The proof (outlined in detail in the appendix) follows easily from the stationary distribution. Using scale parameter properties we can obtain the effect of the parameter on the FWHM, the height and the cumulative probability distribution. We start by noting that since the scale parameter serves only to change the scale of the distribution we must in general have a relationship of proportionality
\begin{equation}
    x_{m} \sim \beta,
\end{equation}
which is indeed what we observe for both $D$ and $a$ comparing Eqs. (\ref{max scale}) and (\ref{beta}). We note that since $m$ for example is not a scale parameter we do not expect the same proportionality relationship to hold for varying $m$.

Analogously the locations of the FWHM $x_{\pm H}$ compared to $x_{m}$ are clearly
\begin{equation}
    p_s(x_{\pm H};\beta) \sim p_s(x_{\pm H}/\beta; 1) = \frac{1}{2}p_s(x_{m}/\beta; 1).
\end{equation}
and thus
\begin{align}
    \text{FWHM} &= (x_H - x_{-H})_1 \nonumber \\
    &= \left(\beta x_H - \beta x_{-H}\right)_\beta \sim \beta,  \label{FWHM scale}
\end{align}
where $(\cdot)_\beta$ represents consideration of the distribution $p_s(\cdot; \beta)$. The cumulative distribution $F(x) = \int^x p(x')\,dx'$ is affected by the scale parameter according to
\begin{equation}
    F(x;\beta) = F(x/\beta;1).
\end{equation}

\subsection{Cases}
In the following we consider $n, m \geq 0$. We define $\gamma = 1/(n-2m+1)$. 

The regime $n - 2m < -1$ ($\gamma < 0$) is shown in Fig. \ref{m03}(a)-(b). This distribution does exist and is normalizable, however has diverging moments and a diverging tail (see appendix). This regime is characterized by a well defined finite maximum and FWHM, as well as on-off intermittency. In this regime increasing $D$ surprisingly causes the distribution to become more concentrated around the noise induced maximum both in the sense of the maximum $\sim D^{-|\gamma|}$ and the FWHM $\sim D^{-|\gamma|}$ of Eqs. (\ref{max scale}) and (\ref{FWHM scale}). It could be said that the noise has a more attracting effect. In fact (as shown in the appendix) as $D \rightarrow \infty$ the stationary distribution surprisingly converges to a delta function. This regime will be the focus of the remaining sections of this paper. We first briefly discuss the two other regimes.

In the case of $n - 2m > -1$ (or $\gamma > 0$) it is clear from Eq.~(\ref{beta}) that increasing $D$ causes the stationary distribution to become more spread out over the non-confining potential  in the sense of $x_m \sim D^{\gamma}$ and FWHM $\sim D^{\gamma}$ (see Fig. \ref{m03}(e)-(f)). $D$ has a more typical diffusion effect. In this case the extremum of the distribution, which can be obtained from Eq.~(\ref{max}) is a minimum. As obtained in Eq.~(\ref{self_sim_norm}) the normalization constant becomes infinite and the particle consequently has probability zero of being in any particular location in the infinite time limit. The distribution is clearly non-stationary. 

The boundary between the two discussed cases is at $n - 2m = -1$ ($\gamma = 0$). It is shown in the appendix that in Itô as well as in Stratonovich interpretation the noise induces a drift term $g(x)g'(x)$, which may counter act the drift term $f(x)$. The equality holds if and only if the deterministic drift term and the stochastic drift term have the same dependency on $x$, i.e. $f(x) \sim g(x)g'(x)$. This means that no maxima/minima can exist as can be seen from Eq.~(\ref{max}). There is no maximum and no scale parameter. The distribution is in that sense 'scale' free. It has been shown by Kaulakys et al. \cite{kaulakys2006nonlinear, kaulakys2009modeling} that such SDEs (with $n - 2m = -1$) correspond to $1/f$ noise. This noise has the stationary distribution $p_{s}(x) \sim 1/x^\lambda$ and spectral density $S(f) \sim 1/f^\alpha$ where $\alpha = (4m - 5 - a/D)/(2m - 2)$.

\subsection{Stationarity of $n - 2m < -1$}
Is the regime $n - 2m < -1$ of Eq.~(\ref{sde_monomial}) stationary? A (strict-sense) stationary process is one whose density function is invariant to time shifts. Typically this is taken to mean that the moments are all finite and are time independent. A wide-sense stationary process is one where first and second order properties are finite and independent of time. Generally this is understood to mean that the first two moments $\langle x(t) \rangle$ and $\langle x(t + \tau)x(t) \rangle$ are independent of time $t$.

In the regime $n - 2m < -1$ of Eq.~(\ref{sde_monomial}) all of the moments $\langle x^l \rangle$ with $l = 1, 2, ..$ diverge to infinity. This is clear in Itô formalism since the evolution of the mean is the same a that of the deterministic system, which clearly diverges for $\dot{x} = ax^n$ with $a > 0$ and $n \geq 0$ (and from this follows also the divergence of the higher moments). Thus, in this sense (for $a > 0$) the distribution is not stationary, even in a wide sense. The distribution does however exist in this regime, as evidenced by Eq. (\ref{stat dist}). For this reason we argue the moments are inappropriate to characterize the system of Eq. (\ref{sde_monomial}). If instead the first and second order properties are taken to be the maximum and FWHM then this distribution is in fact stationary in a wide sense. We will refer to this kind of wide-sense stationarity as M-wide sense stationary (meaning wide-sense stationary in terms of its maximum and FWHM). The distribution of Eq. (\ref{sde_monomial}) when $n - 2m < -1$ is M-wide-sense stationary, but not wide-sense stationary in the sense of moments. Let us now explore this type of stationarity in some more detail. 

\begin{figure}
	\begin{center}
		\hspace*{-1cm}\includegraphics[width=9cm]{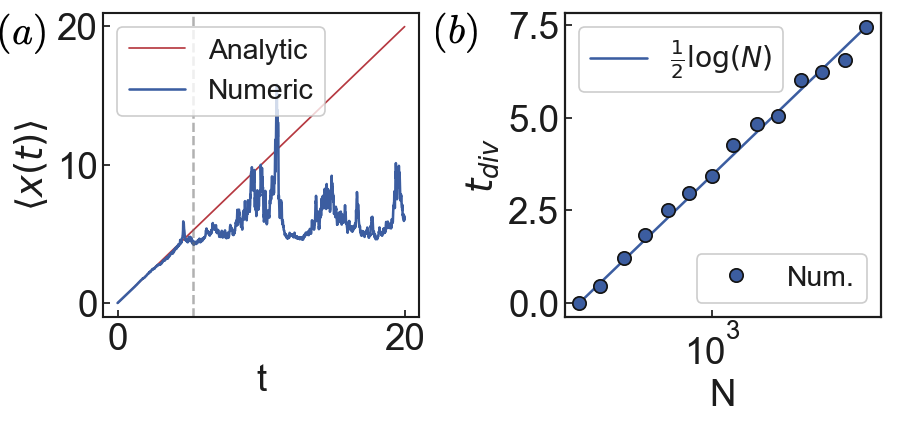}\caption{(a) Time plot of Eq.~(\ref{sde_monomial}) averaged over $ = 10^6$ trajectories, $a = 1, n = 0, m = 1, D = 2$ and $dt=10^{-4}$. Red line shows theoretical mean value. (b): Divergence point of theoretical and measured mean (grey line in (a)).}\label{mean2}
	\end{center}
\end{figure}

The trajectories of the SDE of Eq. (\ref{sde_monomial}) when $n - 2m < -1$ exhibit bursts (see Fig. \ref{m03}). If we observe a single trajectory, such as that of Fig. \ref{m03}(a) we see that regardless of initial conditions the particle will quickly return to the vicinity of the maximum. Due to this, regardless of the initial condition, the correct maximum and FWHM can be established even for relatively short trajectories. This does not change over time. The mean on the other hand cannot in general be established for long trajectories (see Fig. \ref{mean2}). 
Due to the heavy tail of the probability distribution the larger bursts dominate the smaller bursts in size by orders of magnitude (Fig. \ref{mean2}(a)). Since the probability of such a burst having happened increases over time, this means that for a finite ensemble of trajectories the probability of the bursting distorting the mean increases over time. This means that larger sample sizes are increasingly necessary to determine the mean accurately as time evolves. The fact that the trajectory keeps returning to the vicinity of the maximum is not in contradiction to the fact that the mean continually increases. This apparent contradiction can be explained by the fact that as time evolves, the probability of having had a large burst increases.

To see why it is most appropriate to consider the distribution in terms of its maximum and FWHM instead of its moments, we run a large ensemble of trajectories (up to $N = 10^7$) of Eq. (\ref{sde_monomial}) with initial conditions at $x_0 = 0$ (Fig. \ref{mean2}(b)). We see clearly that the average of a small ensemble may represent the theoretical mean faithfully only up to a certain time $t_{\text{div}}$, after which the two values diverge.

This amount of time $t_{\text{div}}$ can be clearly observed to increase logarithmically with system size $N$. This means that it is unrealistic to describe any more than a fairly short trajectory $t_{\text{div}} \approx 10$ in terms of the mean using only standard computers.

\section{Laminar distribution}
We now look at the statistical properties of the bursting for the on-off intermittency case $n-2m < -1$. The particle has a laminar length, defined as the amount of time (or length $l$) between two bursts. A burst is defined as quick event where the trajectory exceeds a certain threshold $x_c$. To calculate this distribution we begin with the time dependent Fokker-Planck equation of Eq. (\ref{sde_monomial}) is
\begin{equation}
    \frac{\partial p(x, t)}{\partial t} = - \frac{\partial}{\partial x}[ax^n p(x, t)] + D\frac{\partial^2}{\partial x^2}[x^{2m} p(x, t)]
\end{equation}
As the coordinate of the system stays for a long time in the region $x < x_c$, one can suppose that the probability density may find the form of a meta-stable distribution decaying for a long period of time. The relaxation process of the probability density to this meta-stable state is supposed to be very fast in comparison with the time of the meta-stable distribution decay, therefore one can neglect the transient $0 \leq t \leq t_{tr}$ \cite{hramov2007length, hirsch1982theory, kye2000characteristic}. Under the assumptions above the probability density may be written as 
\begin{equation}
    p(x, t) = A(t)q(x)
\end{equation}
for small $x$. $q(x)$ can be solved in much the same way as the stationary distribution . The decrease of $A(t)$ should be determined by the probability distribution taken in the boundary point $x_c$, i.e., $dA(t)/dt \sim -p(x_c, t)$. This assumption, which is also equivalent to neglecting the time correlation of the orbit, may be rewritten as
\begin{equation}
    \frac{dA(t)}{dt} = -k A(t) \frac{1}{D x_{c}^{2m}}\exp{\frac{a}{D}\frac{x_{c}^{n-2m+1}}{n-2m+1}}
\end{equation}
where $k$ is a proportionality coefficient. In the absence of a further constraint it may be that both $k$ and $A(t)$ are functions of $D$. Evidently the decrease of $A(t)$ is described by the exponential law.
\begin{equation}
    A(t) = A(0)\exp{-k\eta t}.
\end{equation}
where
\begin{equation}
    \eta = \frac{k}{D x_{c}^{2m}}\exp{\frac{a}{D}\frac{x_{c}^{n-2m+1}}{n-2m+1}},
\end{equation}

\begin{figure}
	\begin{center}
		\hspace*{-1cm}\includegraphics[width=8cm]{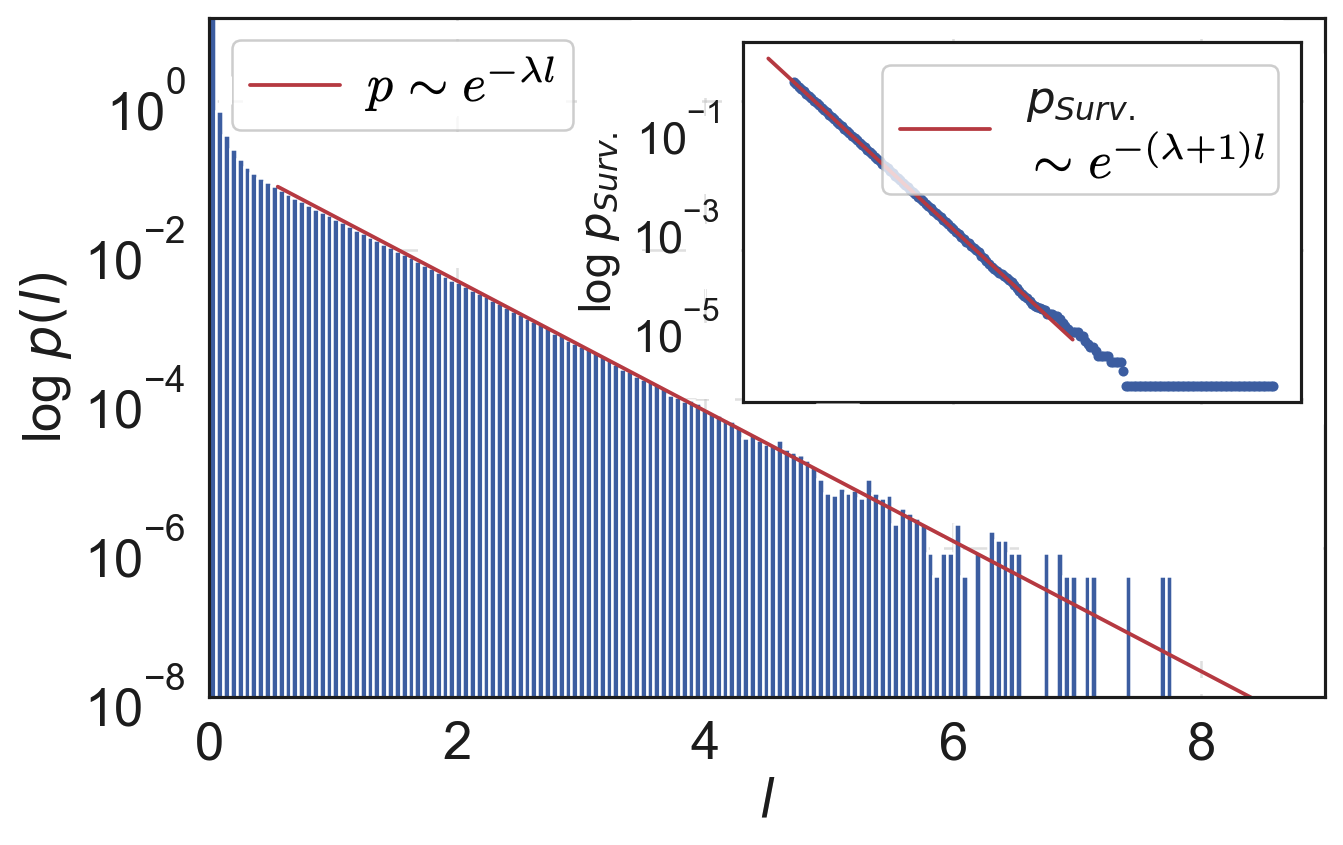}\caption{Stationary probability distribution of the laminar length $l$. Data obtained using single long simulation and waiting for particle to return to region of fixed point. Plotted logarithmically and fitted with exponential fit $p(l) \sim e^{-\lambda l}$. Inset plot is survival probability density over $l$.}\label{poisson 2}
	\end{center}
\end{figure}

This is equivalent to the exponential law for the laminar phase distribution, since this distribution is defined as 
\begin{equation}
    l(t) = -\int_{0}^{\infty} \frac{\partial p(x,t)}{\partial t}\,dx 
\end{equation}
and thus defining $\widetilde{T}^{-1} = \eta k$ we have 
\begin{equation}
    l(t) = \widetilde{T}^{-1}\exp(-t/\widetilde{T})
\end{equation}
This exponential decay of the laminar lengths is shown in Fig. \ref{poisson 2}.

\subsubsection{Mean First Passage Time}
\begin{figure}
	\begin{center}
		\hspace*{-0.3cm}\includegraphics[width=10cm]{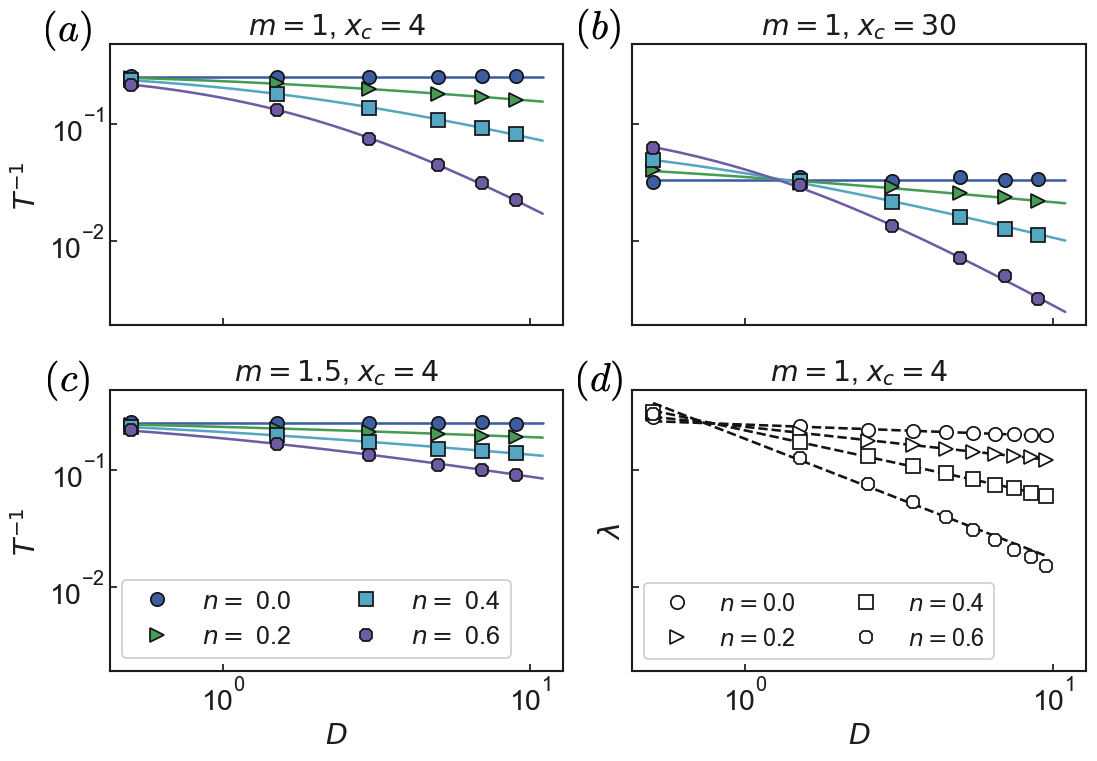}\caption{(a)-(c) Inverse MFPT. Lines are analytical results obtained from Eq.~(\ref{MFPT}) obtained with Simpson's rule, circles are numerics. (d) Circles are $\lambda$ obtained from exponential fit shown in Fig. \ref{poisson 2}. Lines are power law fit.}\label{4plot_mfpt}
	\end{center}
\end{figure}

For the purposes of further theoretical treatment we will explore the Mean First Passage Time (MFPT). The first passage time is defined as the time taken to travel from $x_0$ to the threshold $x_c$. We choose $x_0 = 0.1$. Given that bursts quickly return back to the vicinity of $g(x) = 0$ we expect the distribution of laminar lengths to be equivalent to the distribution of first passage times. The mean of this distribution can be calculated using numerical methods (see appendix) considering reflecting boundaries at $x = 0$ as \cite{gardiner2009stochastic}
\begin{equation}
    T(x_0) = \frac{1}{D}\int_{x_0}^{x_c}\frac{dy}{\psi(y)}\int_{0}^{y}\frac{\psi(z)}{z^{2m}}\,dz \label{MFPT}
\end{equation} 
where we have defined $\psi(x) \vcentcolon= \exp{\frac{a}{D}\frac{x^{n-2m+1}}{n-2m+1}}$. A plot of the Mean First Passage Time for different $m$ is shown in Fig. \ref{4plot_mfpt}(a)-(c). We see that a power law becomes an increasingly good approximation of the MFPT as the threshold boundary $x_c$ is increased (compare Figs. \ref{4plot_mfpt}(a) and (c)). We see that (for $n > 0$) the frequency of the bursts $T^{-1}$ (equivalent to $l$) decreases with $D$. Increasing the multiplicative noise intensity thus stabilizes the system also in the sense of longer laminar lengths (in addition to in the sense of the maximum and FWHM). We see by comparing Figs. \ref{4plot_mfpt}(a) and (c) that multiplicative noise $D$ has less of an effect for larger $m$. This is presumably due to the trajectory being drawn closer to the minimum $g(x) = 0$ for larger $m$ where the noise term $\sqrt{2D}x^m$ is smaller.

In Fig. \ref{4plot_mfpt}(d) numerical results obtained from the decay constant $\lambda$ of the exponential distribution are shown. These are obtained by measuring a survival probability density determined in stochastic simulations (see Fig. \ref{poisson 2}). In calculating this density we do not consider laminar lengths under a given threshold $l_0 < 0.28$, which are very sensitive to the choice of initial conditions and as such do not obey an exponential distribution. In Fig. \ref{4plot_mfpt}(d) it is clearly visible that $\lambda$ decays with $D_m$ according to a power law $\lambda \propto D_m^{-\alpha}$. The agreement between the inverse MFPT $T^{-1}$ and $\lambda$ as $x_c$ becomes large is to be expected, since the higher the upper boundary the less sensitive is $T^{-1}$ to the initial conditions.

\section{Generalizations}
\subsection{Generalized drift and Stratonovich interpretation}

\begin{figure*}
	\begin{center} 
		\hspace*{-0.5cm}\includegraphics[width=17.5cm] {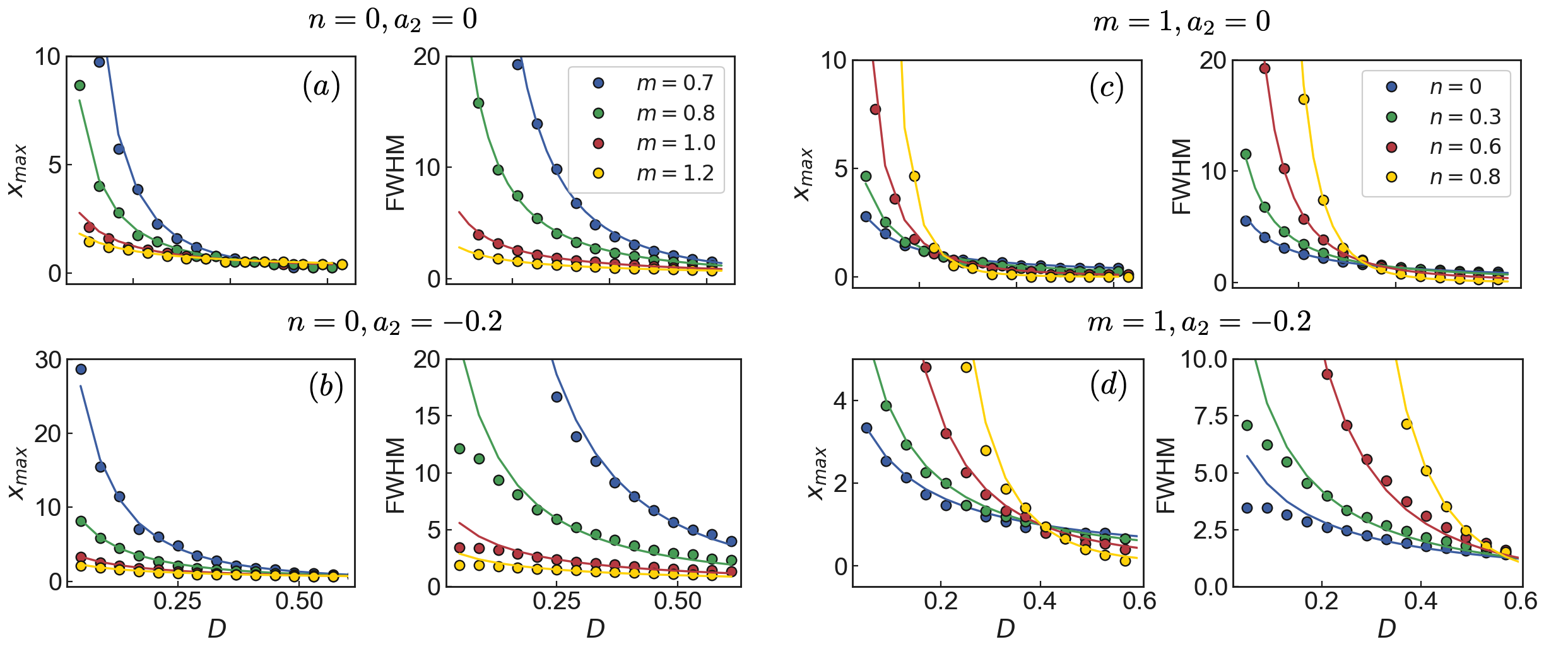}\caption{$x_{max}$ and FWHM of Eq. (\ref{sde w a2}). (a) for fixed $n = 0, a_2 = 0$ ($a_1 = 0.5$), (b) $n = 0, a_2 = -0.2$ ($a_1 = 1$), (c) $m = 0, a_2 = 0$ ($a_1 = 0.5$) and (d) $m = 0, a_2 = -0.2$ ($a_1 = 1$). Lines of $x_{max}$ and FWHM are fits using Eqs. (\ref{max scale}) and (\ref{FWHM_gen}) respectively.}\label{fwhm_and_max}
	\end{center}
\end{figure*}

The first generalization is the equation 
\begin{equation}
    \dot{x} = a_{1}x^n + a_{2}x^{2m-1} + \sqrt{2D}x^m \xi(t). \label{sde w a2}
\end{equation}
The maximum of this equation can be easily obtained from the derivative of the stationary probability density as 
\begin{equation}
    x_m = \left(\frac{(2-\nu)mD - a_2}{a_1} \right)^\frac{1}{n-2m+1}.  \label{eq. max}
\end{equation}
We see that (assuming $a_1 > 0$) the maximum is only well defined if $a_2 < (2-\nu)mD$ as well as $n - 2m + 1 < 0$. Eq. (\ref{sde w a2}) can using the following substitution
\begin{equation}
    y(x) = \alpha^{-1}x^{\alpha} ~~\text{where}~~ \alpha = 1 - (a_{2}/D) \label{subs}
\end{equation}
and Itô's lemma \cite{gardiner2009stochastic} be brought into the original form
\begin{equation}
    \dot{y} = a_{1}(\alpha y)^{\frac{n+\alpha-1}{\alpha}} + \sqrt{2D}(\alpha y)^{\frac{m+\alpha-1}{\alpha}}\xi(t).  \label{gen drift}
\end{equation}
Since $y(x)$ is proportional to $x^\alpha$ we can see that from M-stability of $y$ must also follow M-stability of $x$. Considering the exponents of Eq.~(\ref{gen drift}) instead of those of Eq.~(\ref{sde_monomial}) we see that the $\alpha$ terms cancel out and that we have M-stability for $n - 2m + 1 \leq 0$. 

We note that if we instead interpret Eq.~(\ref{sde_monomial}) using Stratonovich interpretation then we may rewrite this in Itô formulation by considering the Stratonovich drift term
\begin{equation}
    \begin{split}
        \dot{x} &= a_{1}x^n + a_{2}x^{2m-1} + \sqrt{2D}x^{m}\circ \xi(t)  \\
            &= a_{1}x^n + (a_{2} + mD)x^{2m-1} + \sqrt{2D}x^{m} \xi(t).
    \end{split}
\end{equation}
We see from this and the previous discussion about substitution and use of scale parameters that the FWHM is well defined and stable for this equation as long as $a_2 < mD$ and $n - 2m + 1 < 0$. In other words if Eq. (\ref{sde w a2}) is M-stable in Itô interpretation then it is also M-stable in Stratonovich interpretation as long as the condition $a_2 < (2 - \nu)mD$ holds. We have seen that FWHM $\sim (1/a_{1})^{\frac{1}{n-2m+1}}$. 

It is interesting to ask more generally whether $D$ behaves as a scale parameter for the more general Eq. (\ref{sde w a2}) (where $a_2$ is a constant independent of $D$)? In this case we would expect the FWHM to vary with $D$ as the maximum does according to
\begin{equation}
    \text{FWHM} \sim \left(\frac{(2-\nu)mD - a_2}{a_1} \right)^\frac{1}{n-2m+1}.  \label{FWHM_gen}
\end{equation}
This claim is explored in Figs. \ref{fwhm_and_max}(a)-(b) for different $m$ and Figs. \ref{fwhm_and_max}(c)-(d) for different $n$. We see for $a_2 < 0$ in (b) and (d) that agreement between the conjecture and numerical results agrees increasingly well as $D$ becomes larger. This is to be expected, since $\alpha \rightarrow 1$ and thus $y(x) \rightarrow x$ as $D \rightarrow \infty$. For $a_2 > 0$ the agreement seems to be good even for smaller $D$ (Results shown in the appendix). Numerical simulations suggest that $a_2$ is not a scale parameter.

\begin{figure*}
	\begin{center}
		\hspace*{-1cm}\includegraphics[width=15cm]{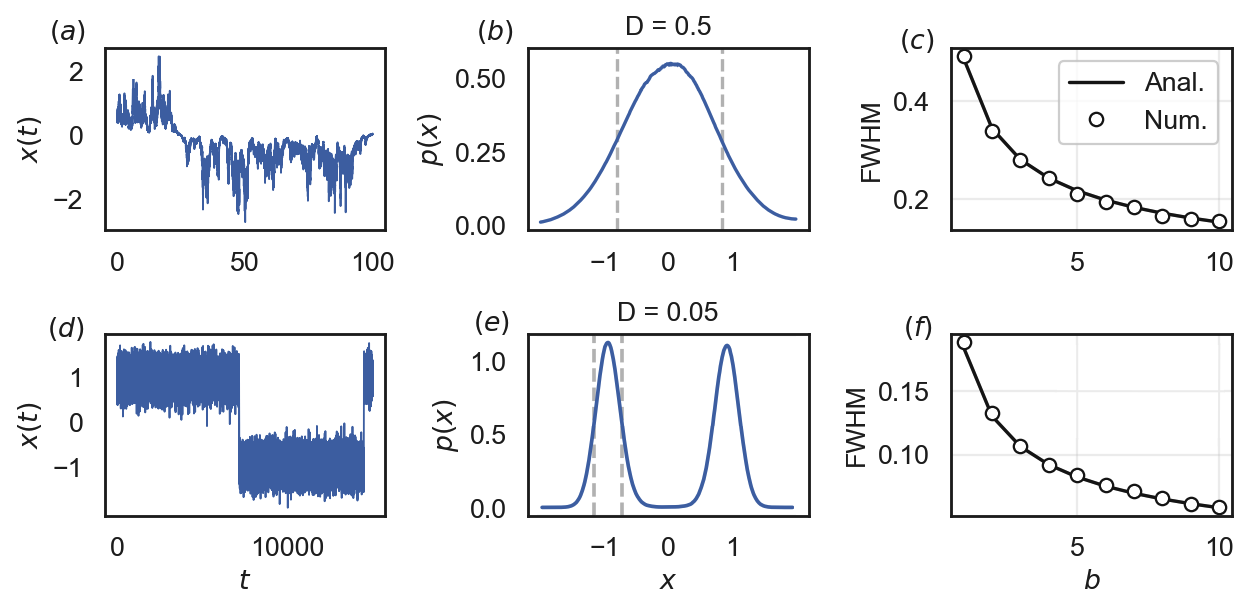}\caption{Double well potential of Eq. (\ref{double well}) with parameters $a = 1, b = 1$, (a)-(c) $D = 0.5$, (a) Time plot, (b) stationary probability distribution, (c) FWHM of (in deterministic case unstable) fixed point. (d)-(f) same as (a)-(c) with $D = 0.05$, whereby a small amount of additive noise $D_{add} = 0.01$ has been added to allow jumping across $x = 0$.}\label{double_well_fwhm_unst}
	\end{center}
\end{figure*}

\subsection{Additive noise}
We now consider the more general case where additive noise $\xi_{\text{add}}$ is added to the SDE, i.e. we write Eq. (\ref{SDE}) as
\begin{align}
\begin{split}
    \dot{x} &= f(x) + \sqrt{2D_{\text{m}}}g(x)\xi_{\text{m}}(t) + \sqrt{2D_{\text{add}}}\xi_{\text{add}}(t) \\
            &= f(x) + \sqrt{2D_{\text{m}} g(x) + 2D_{\text{add}}}\xi(t) \\
            &= f(x) + \sqrt{2D_{\text{m}}}\widetilde{g}(x)\xi(t),
\end{split}
\end{align}
where $\widetilde{g}(x) \vcentcolon= \sqrt{g(x)^2 + D_{\text{add}}/D_{\text{m}}}$. In the second line we have used the central limit theorem and the fact that the noise is Gaussian distributed. It can easily be shown that $\widetilde{g}(x)\widetilde{g}'(x) = g(x)g'(x)$. For this reason this alteration has no affect on the maximum of the distribution, derived from Eq. (\ref{max}). We therefore expect additive noise to also have no effect on the scale parameter, since $\beta \sim x_m$. We would for example expect the FWHM to vary with $D_{\text{m}}$ in the presence of additive noise just as FWHM varies with $D$ when no additive noise is present. This is indeed the case and will be explored more fully in a follow up paper. 

\section{Example}
In this section we apply the results of this paper to elucidate the role played by multiplicative noise in the tipping points in a double well potential. A classic model containing a double well potential (and multiplicative noise) is that of the (noisy) Duffing oscillator described by the equation \cite{aumaitre2007noise, bourret1973brownian, lindenberg1981brownian}
\begin{equation}
    \frac{1}{\gamma}\ddot{x} + \dot{x} = (a+y(t))x - bx^3,
\end{equation}
where $b$ is the (positive) friction coefficient and $d$ is the mean frequency. $y(t)$ is Ornstein-Uhlenbeck noise with correlation function $\langle y(t)y(0) \rangle = q e^{-\Delta \omega |t|}$. This equation may for example be used to model the stability properties of the conductive state for Rayleigh Bénard convection, if we consider the Rayleigh number to be a noisy function of time \cite{ahlers1984externally, lucke1985response}. In the limits (i) $\gamma \gg \Delta \omega$ and (ii) $\Delta \omega \gg \gamma$ may both be represented by the same formal equation 
\begin{equation}
    \dot{x} = (a + \sqrt{2D} \xi(t))x - bx^3 \label{double well}
\end{equation}
if this equation is interpreted in the sense of Stratonovich in case (i) and in the sense of Itô in case (ii) \cite{graham1982stabilization}. This is the equation of a double well potential with multiplicative noise. In the absence of noise this double well potential has an unstable fixed point at $x = 0$ and two stable fixed points at $x = \pm \sqrt{a/b}$. The stationary probability distribution can be calculated as 
\begin{equation}
    p_{s}(x) = \frac{N}{2D}x^{\frac{a}{D}-2}\exp{-\frac{b}{2D}x^2}.
\end{equation}
The FWHM is not simple to obtain from this equation, since by setting $p_{s}(x_H) = p_{s}(x_m)/2$ and calculating the maximum $x_m$ we obtain the $x_H$ at the point of the FWHM is a function of the non-analytic Lambert W function. Nevertheless, considering the ideas of this paper (comparing $(b/2D)x^2$ and $(x/\beta)^2$) we see immediately that we have a scale parameter $\beta \sim b^{-1/2}$. Thus $\text{FWHM} \sim \beta \sim b^{-1/2}$. This agreement is illustrated in Fig. \ref{double_well_fwhm_unst}.

We now explore the behavior of the system in the vicinity of these fixed points. In the vicinity of $x = 0$ we argue that noise may stabilize the system. To show this we first calculate the maximum of the stationary probability distribution using Eq. (\ref{max}) as $x = \pm \sqrt{(a-2D)/b}$. As $D$ is made larger ($\rightarrow a/2$) the maximum moves closer toward $x = 0$. In this region we make a simplifying approximation of $-bx^3 + ax \approx ax^{1-\epsilon}$ where $\epsilon > 0$ is a small number and $\epsilon \rightarrow 0$ as $D \rightarrow a/2$. We can see immediately that $n = 1-\epsilon$ and $m = 1$ and thus $n - 2m < -1$. In other words the noise makes the fixed point at $x = 0$ M-stable. For $D > a/2$ the maximum of the stationary distribution is at $x = 0$. This again demonstrates the confining role of multiplicative noise, showing that white noise may stabilize the position of the potential well. 

If $D$ is small then we may linearize around $x = \sqrt{a/b}$ to obtain 
\begin{equation}
    \dot{x} \approx 2a\sqrt{a/b} - 2ax + \sqrt{2D}x\xi(t).
\end{equation}
Using the substitution $y(x) = \alpha^{-1}x^\alpha$ where $\alpha = (D + 2a)/D$ we obtain
\begin{equation}
    \dot{y} = 2a\sqrt{a/b}\left(\frac{D+2a}{D}y \right)^{\frac{2a}{D+2a}} + \sqrt{2D}\left(\frac{D+2a}{D}y\right)\xi(t).
\end{equation}
We see that $n = 0$ and $m = 1$ and thus $n - 2m = -2 < -1$. Thus these minima are also M-stable. We also see as noted previously that $b^{-1/2}$ is a scale parameter. While the fact that noise of a similar double well potential system causes the distribution to move toward zero at large intensities has been known from work from Graham et al \cite{graham1982stabilization} that work focused on moments, such as $\langle x^2 \rangle$ related to the variance. It was as such not able to account for the effect of the parameters on narrowing the peaks of the distribution.

\section{Conclusions}
We have shown that under quite general conditions (when $n - 2m < -1$) increasing multiplicative noise intensity shifts the mass of the stationary distribution closer to the minimum of the absolute value of $g(x)$ and behaves as a scale parameter. We have shown that in this range there is exponentially distributed on-off intermittency, and that this goes hand in hand with the distribution having a scale parameter (a kind of self similarity). The fact that both of these phenomena are connected is well known in turbulence. The boundary case $n - 2m = -1$ is scale free power law noise (with a power law spectral density).  

When exploring the analytically most simple case of natural boundaries we have defined the concept of a distribution being $M$-wide sense stationary (having constant maximum and FWHM) and shown that the case of $n - 2m < -1$, even though it is not wide sense stationary in the traditional sense (having diverging moments). We have shown importantly that the FWHM generally varies with the scale parameters in this regime in the same way as the maximum. The maximum and FWHM act as analogues to the first and second moments and collectively describe the body of the distribution, excluding the heavy tail. The bursts have been considered using first passage time theory.

Given essentially only the requirement of a significant multiplicative noise term we expect these results to be quite applicable to a range of different problems including extreme bursts, tipping points and synchronization. We have applied the ideas of this paper to a double well potential as a specific example. In particular these methods are important in the regime of large noise where standard small noise perturbation methods are insufficient. In this regime we have shown that it is sufficient to study the first term of a Taylor expansion of both the deterministic and stochastic terms of the Langevin equation. 

\section{conflict of interest}
The authors have no conflict of interest to disclose.

\section{Appendix}

\subsection{Some basic analytics}

\subsubsection{Delta function as $D \rightarrow \infty$}
In this subsection we show that the stationary distribution of Eq. (\ref{sde_monomial}) with $n - 2m < -1$ becomes a delta function as $D \rightarrow \infty$. To prove these we calculate the integral of this function against a sufficiently good test function $\phi$ as $D \rightarrow \infty$. Rigorously, the Dirac function is only defined when acting on functions in the Schwartz space. Let $\phi$ be an infinitely smooth function on $\mathbb{R}$ with compact support $\phi \in C_{c}^{\infty}(\mathbb{R})$. We define 
\begin{equation}
    f_D \vcentcolon= \frac{N}{x^{2m}}\exp{-\frac{a}{Db^2}\frac{x^{n-2m+1}}{n-2m+1}}.
\end{equation} 
We now calculate the integral of this against our test function (using notation from functional analysis)
\begin{align}
    \left<f_D, \phi \right> &= \int_{0}^{\infty}\frac{N}{x^{2m}}\exp{-\frac{a}{Db^2}\frac{x^{n-2m+1}}{n-2m+1}} \phi(x) \,dx \nonumber \\
    = \int_{0}^{\infty}&\frac{\widetilde{N}}{y^{2m}}\exp{-\frac{a}{b^2}\frac{y^{n-2m+1}}{n-2m+1}} \phi \left(yD^{\frac{1}{n-2m+1}}\right) \,dy
\end{align}
where we have used the fact the $p(x < 0) = 0$ and the substitution $y = xD^{-\frac{1}{n-2m+1}}$. We have defined $\widetilde{N} \vcentcolon= N D^{\frac{1-2m}{n-2m+1}}$. We now look at the case where $n - 2m < -1$. As $\phi$ is continuous and compactly supported, this integrand is dominated by $(\widetilde{N}/y^{2m})\exp(-\frac{a}{b^2}\frac{y^{n-2m+1}}{n-2m+1})\| \phi \|_{\infty}$, which integrates to $\| \phi \|_{\infty}$. Moreover, because $\phi$ is continuous the integrand converges pointwise to $\frac{\widetilde{N}}{y^{2m}}\exp{-\frac{a}{b^2}\frac{y^{n-2m+1}}{n-2m+1}} \phi(0)$ as $D \rightarrow \infty$. Applying the dominated convergences theorem yields
\begin{align}
    \lim_{D \rightarrow \infty}\left<f_D, \phi \right> &= \frac{\widetilde{N}}{y^{2m}}\exp{-\frac{a}{b^2}\frac{y^{n-2m+1}}{n-2m+1}} \phi(0) \nonumber \\ &= \phi(0)
\end{align}
since the stationary probability distribution is normalized to one. Thus
\begin{equation}
    \lim_{D \rightarrow \infty}\left<f_D, \phi \right> = \left<\delta_0, \phi \right>.
\end{equation}
This result is in a way very counter intuitive. As additive noise intensity increases distributions become wider and flatter (diffusion). Multiplicative noise in this sense seems to have exactly the opposite effect as additive noise, causing the distributions to become increasingly more peaked. 

\subsubsection{Generalized minima of $g(x)$}
\begin{figure}
	\begin{center}
		\hspace*{-0.4cm}\includegraphics[width=8.5cm]{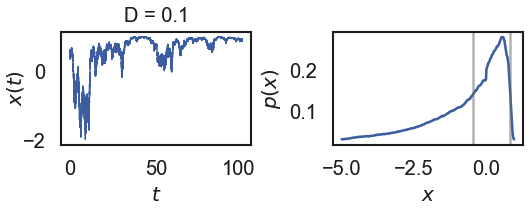}\caption{Time plot and stationary solution of FPE of Eq. (\ref{eq_bx1}). Grey lines indicate FWHM.}\label{bound_x_1}
	\end{center}
\end{figure}

If $g(x)$ has a different zero $g(x_u) = 0$ to $x = 0$ or if the deterministic drift is instead going in the negative direction and the noise creates a gradient in the positive direction such as 
\begin{equation}
    \dot{x} = -a|x|^{n} + \sqrt{2D}c|x_u - x|^m \xi(t),
\end{equation}
then we may use a transformation $y = x_u - x$ to bring the equation into the form 
\begin{equation}
    \dot{y} = a(y - x_u)^{n} + \sqrt{2D}y^m \xi(t),
\end{equation}
which in the case of $n = 1$ takes the form of Eq.~(\ref{sde_monomial}). In the case that the drift and diffusion are acting in the same direction the variable will simply drift into the zero $g(x) = 0$ at which point the noise will switch off. An example is shown in Fig. \ref{bound_x_1} for the equation 
\begin{equation}
    \dot{x} = -ax^n + b(x-1)^m \xi(t)  \label{eq_bx1}.
\end{equation}
The trajectory has laminar lengths in the vicinity of $x = 1$ and experiences bursts in the negative direction due to the negative drift term. 

\subsubsection{Noise induced drift}
In this section we discuss the drift induced by multiplicative noise. In the Stratonovich case noise induces a well known 'Stratonovich' drift $Dg(x)g'(x)$. In the Itô interpretation this drift term is also present. To justify that claim we use a substitution, which obeys $h'(x) \vcentcolon= \frac{1}{g(x)}$. Thus the general substitution must be $y = h(x) = \int \frac{dx}{g(x)}$ and $h''(x) = -g'(x)/g^{2}(x)$. In this way we obtain (the well known equation)
\begin{align}
    dy &= h'(x)\,dx + \frac{1}{2}h''(x)\,dx^2 \nonumber \\
    &= \frac{1}{g(x)}\left[f(x) - Dg(x)g'(x) \right]\,dt + \sqrt{2D}dW_t.
\end{align}
Due to the symmetry of the white noise the function $g(x)$ can be replaced with the function $|g(x)|$. In this way it can be seen that $y = h(x)$ is a monotonic function of $x$. This is in line with the behaviour of the maximum (Eq. (\ref{max})).

It is clear that when $g(x)g'(x)$ varies more strongly $x$ over the domain than $f(x)$ then the overall drift direction of the particle (at large $x$) is toward the zero $g(x) = 0$ (this is a generalization of the case $n - 2m < -1$). On the other hand if $f(x)$ varies more strongly $x$ over the domain than $g(x)g'(x)$ then the dominant force on the particle (at large $x$) is forcing it toward the fixed point $f(x) = 0$ (this is a generalization of $n - 2m > -1$). 

\subsubsection{FPE}
The following theory is well known (see for example \cite{gardiner2009stochastic, horsthemke1984noise}). The Fokker Planck Equation is given by 
\begin{align}
\begin{split}
    \frac{\partial}{\partial t}p(\phi, t) &= -\frac{\partial}{\partial \phi}[\{f(\phi) + \frac{\nu}{2}g'(\phi)g(\phi)\} p(\phi, t)] \\
    &~~~~+ \frac{1}{2}\frac{\partial^2}{\partial \phi^2}[g^2(\phi)p(\phi, t)] 
\end{split} \\
    &= -\frac{\partial}{\partial \phi}J(\phi, t)
\end{align}
We wish to obtain the stationary distribution $\partial_t p_{s}(\phi, t) = 0$. The solution to the homogeneous equation, known as the fundamental solution is found to be 
\begin{align}
     \psi(\phi) &\vcentcolon= \exp[2\int^{\phi}\frac{f(\phi') + \frac{1}{2}g'(\phi')g(\phi')}{g^2(\phi')}\,d\phi'] 
\end{align}
The complete solution is found to be 
\begin{equation}
    p_s(\phi) = \frac{\psi(\phi)}{g^2(\phi)}\left(N - 2J \int^{\phi}\frac{d\phi'}{\psi(\phi')} \right).
\end{equation}
In the case of absorbing or natural boundary conditions $J=0$. Thus the stationary solution becomes
\begin{equation}
    p_{s}(x) = \frac{N}{g^{2}(x)}\exp{2\int^{x}\frac{f(u)}{g^{2}(u)}du} \label{stat_dis_abs}
\end{equation}

\subsubsection{Proof of scale parameter}
To show this we prove that $p_s(x;\beta) \sim p_s(x/\beta; 1)$. From this Eq.~(\ref{scaling dist 2}) follows automatically, due to normalization. We now prove the first statement. Since $g(x)g'(x) = 2Dmx^{2m-1}$, the left hand side of Eq.~(\ref{scaling dist 2}) is equal to
\begin{align}
    p_{s}(x;\beta) &= \frac{N}{2D g^{2}(x)}e^{\frac{1}{D}\int_{A}^{x}\frac{f(u)+(\nu/2)g'(u)g(u)}{g^2(u)}\,du} \nonumber \\
    &\sim x^{(\nu-2)m}\exp{\frac{a}{D}\frac{x^{n-2m+1}}{n-2m+1}},
\end{align}
where $A$ is the left boundary, and $\nu = 1$ in Stratonovich interpretation and $\nu = 0$ in Itô interpretation. On the other hand we also have
\begin{align}
    p_{s}(x/\beta;1) &= N(x/\beta)^{-2m}e^{\int^{x/\beta}\frac{u^n}{u^{2m}}+\frac{m}{u}\,du} \nonumber \\ 
    &\sim x^{(\nu-2)m}\exp{\frac{(x/\beta)^{n-2m+1}}{n-2m+1}}.
\end{align}
From simple comparison $\beta$ is obtained. Thus Eqs.~(\ref{scaling dist 2}) and (\ref{beta}) have been proven. The normalization constant is calculated as
\begin{align}
    N^{-1} &= \frac{1}{2D}\int_{0}^{\infty}x^{(\nu-2)m}e^{\frac{a}{D}\frac{x^{n-2m+1}}{n-2m+1}} \nonumber \\
\begin{split}
    &= \frac{1}{2a}x^{\nu m - n}\left(\frac{a}{D\delta}x^{-\delta}\right)^{\frac{\nu m - n}{\delta}} \\ &~~~~~\cdot \left. \Gamma \left(\frac{(2-\nu)m-1}{\delta},\frac{a}{D\delta}x^{-\delta}\right) \right\vert_{0}^{\infty}  \label{self_sim_norm} 
\end{split}
\end{align}
where $\delta \vcentcolon= 2m-n-1 > 0$. In Itô interpretation $N^{-1} < \infty$ for all $\delta > 0$. In Stratonovich interpretation this is only guaranteed if we impose the additional requirement that $m > 1$. In the case that $n = 0$ and $m = 1$ in Stratonovich interpretation we have 
\begin{equation}
    N^{-1} \sim \lim_{b\to \infty}\int_{0}^{b}\frac{e^{-\frac{a}{Dx}}}{x}\,dx  = \Gamma(0),
\end{equation}
which is undefined. A similar result holds for all $m < 1$. In the special case $n = 0$ and $m > 0.5$ in Itô interpretation we obtain the simple result
\begin{equation}
    N^{-1} = \frac{1}{2D}\int_{0}^{\infty} x^{-2m}\exp{\frac{a}{D}\frac{x^{1-2m}}{1-2m}}\,dx = \frac{1}{2a}
\end{equation}
In fact more generally we notice in Itô interpretation for $m = 1$ ($n < 1$ follows from $\delta > 0$) that 
\begin{equation}
    N^{-1} \sim \frac{1}{D}\left(\frac{D}{a}\right)^{-\frac{1}{n-1}}.
\end{equation}
The fact that for $P_{s}(x)$ is normalizable when $m \leq 1$ only in Itô interpretation, but not in Stratonovich interpretation requires some discussion. It is generally thought that both Itô and Stratonovich interpretation should lead to qualitatively similar results. It is suggested in \cite{horsthemke1984noise} (page 112) that if an SDE admits a stationary solution in Itô interpretation, but not in Stratonovich interpretation (or vice versa), then such a discrepancy should be interpreted as a "red warning light", and may indicate pathological features. 

Our model does not seem to have any pathological features in its domain $x \in (0, \infty)$ such as not being differentiable at a point. Beyond this, in this paper we are interested in the maximum and FWHM, which are both independent of the normalization and are in fact qualitatively similar. Both distributions are M-stable. For this reason we are not concerned with the fact that one is normalizable while the other is not, which is a reflection only of the heaviness of the tails. The tails are however nevertheless also qualitatively the same, both decaying according to a power law. 

We now ask in the more general case of Eq. (\ref{sde w a2}) if there exists a scale parameter analogous to $D^{\frac{1}{n-2m+1}}$ when $a_2 \neq 0$?
\begin{align}
    p_{s}\left(x; \beta \right) &= \frac{N}{2D}x^{\frac{a_2}{D} - 2m}e^{\frac{a_1}{D}\frac{x^{n-2m+1}}{n - 2m + 1}} \nonumber \\
    &\sim x^{\frac{a_2}{D} - 2m}e^{\frac{a_1}{D}\frac{x^{n-2m+1}}{n - 2m + 1}} \label{stat dist: gen} \\
    p_{s}\left(x/ \beta; 1 \right) &\sim (x/\beta)^{\frac{a_2}{1} - 2m}e^{\frac{(x/\beta)^{n-2m+1}}{n - 2m + 1}} \nonumber \\
    &\sim x^{a_2 - 2m}e^{\frac{(x/\beta)^{n-2m+1}}{n - 2m + 1}}
\end{align}
These results seem to show from the mismatches in the power law terms in general that $D$ is not a scale parameter unless $a_2 = 0$ or $a_2 \propto D$ (equivalent to $a_2 = 0$ in Stratonovich interpretation). $a_1$ is in general a scale parameter. We see that $x$ scales according to $\beta_x \sim \text{FWHM} \sim (1/a_{1})^{\frac{1}{n-2m+1}}$. Although the parameters $a_2$ and $D$ are inextricably linked, and $a_2$ is not a scale parameter, we do still see in the limit of large $D$ or small $a_2$ (more precisely $D \gg a_2$) that $\beta_x \sim D^{\frac{1}{n-2m+1}}$.  

\subsubsection{Diverging moments}
We show diverging moments of Eq. (\ref{stat dist: gen}) when $\delta = 2m - n - 1 > 0$. We first note that due to the average of the noise being zero the expectation of the trajectory is identical to the deterministic trajectory, i.e. $\langle \dot{x} \rangle = f(x)$. Since the potential is chosen to be non-confining the $\langle x \rangle$ drifts to infinity. 

To calculate the moments of the stationary distribution more generally we start with Eq. (\ref{stat dist: gen})
\begin{equation}
    p_{s}(x) = \frac{N}{2D}x^{\frac{a_2}{D} - 2m}\exp{-\frac{a_1}{D}\frac{x^{-\delta}}{\delta}} 
\end{equation}
The first moment is calculated as
\begin{align}
    \langle x \rangle &=  \frac{N}{2D} \int_{0}^{\infty} x^{1 + \frac{a_2}{D} - 2m}\exp{-\frac{a_1}{D}\frac{x^{-\delta}}{\delta}} \,dx \\
    &= \frac{1}{\delta}x^{\nu}\left(\frac{a_1 x^{-\delta}}{D\delta}\right)^{\frac{\nu}{\delta}} \left. \Gamma \left(-\frac{\nu}{\delta}, \frac{a_1}{D\delta}x^{-\delta}\right)  \right\vert_{0}^{\infty}  \\
    &= \infty,
\end{align}
where $\nu \vcentcolon= 2 + (a_2/D) - 2m$. Since the first moment diverges so too do all of the higher moments. 

\subsubsection{MFPT}
We now look at intermittency without resetting. For the case that $D_2 = 0$ we can assume reflecting boundaries as $x = 0$ and the above equation for the MFPT can be simplified dramatically to 
\begin{equation}
    T(x) = \langle T \rangle = \int_{0}^{\infty}\int_{0}^{b}\rho(x',t|x,0)\,dx'\,dt,
\end{equation}
which has a general solution 
\begin{equation}
    T(x) = 2\int_{x}^{b}\frac{dy}{\psi(y)}\int_{a}^{y}\frac{\psi(z)}{g^{2}(z)}\,dz.
\end{equation}
We now look at Eq.~\ref{sde_monomial} ($n = 0$, $m = 1$) and calculate the fundamental solution
\begin{equation}
   \psi(x) = \exp{\int_{a}^{x}dx'\,\frac{\omega}{D x^2}} = C e^{-\frac{\omega}{D x}},
\end{equation}
where $C$ is a constant. When the left hand side has a reflecting boundaries this becomes 
\begin{align}
    T(x) &= 2\int_{x}^{b}\,dy e^{\frac{\omega}{Dy}}\int_{0}^{y} \frac{e^{-\frac{\omega}{Dz}}}{2Dz^2}\,dz \nonumber \\ 
    &= \int_{x}^{b}\frac{dy}{\omega} = \frac{b - x}{\omega}
\end{align}
From this we can obtain the mean laminar length $\langle l \rangle$ as 
\begin{equation}
    \langle l \rangle = | T(b) - T(x) | = \frac{|b - x|}{\omega},
\end{equation}
and thus the number of bursts in a time period $L$, which we denote as $\chi$ is 
\begin{equation}
    \chi(L) = \frac{L}{\langle l \rangle} = \frac{2\pi}{\langle l \rangle}
\end{equation}
It is known that the absolute values of the outfluxes of probability are related to the conditional escape time densities \cite{lindner2001optimal}.

\subsubsection{Additive noise}
The stationary probability distribution is given by
\begin{align}
    p_{s}(x) &= \frac{N}{2D_1 x^{2m} + 2D_2}e^{2\int^{x}\frac{au^n}{2D_1 u^{2m} + 2D_2}\,du} \\
    &= \frac{N}{2D_1 x^{2m} + 2D_2}\cdot \nonumber \\
    &\exp{\frac{ax^{n+1}}{D_2 (n+1)} {}_2 F_1 \left(1, \frac{n+1}{2m}; \frac{\chi}{2m}; \frac{-x^{2m}}{D_2/D_1}\right)}, \nonumber
\end{align}
where $\chi = 2m+n+1$. For the case of $n = 0$ and $m = 1$ this reduces to 
\begin{equation}
\begin{split}
    p_{s}(x) &= \frac{N}{2D_1 x^{2m} + 2D_2}\cdot  \\
    & \cdot \exp{\frac{2a}{\sqrt{D_1 D_2}}\left[n\pi + \tan^{-1}\left(\sqrt{\frac{D_1}{D_2}}x\right)\right]}
\end{split}
\end{equation}
where $n \in \mathbb{Z}$ is defined such that $\pi \left(n-\frac{1}{2} \right) \leq \theta < \pi \left(n+\frac{1}{2} \right)$. Simulations show that the FWHM decreases as $\sim D^{\frac{1}{n-2m+1}}$. We assert that additional additive noise $D_2$ makes no difference to this behavior. To be more explicit the noise does broaden the distribution, making the FWHM wider. The FWHM is nevertheless made smaller again by increasing multiplicative noise according to $\sim D^{\frac{1}{n-2m+1}}$.

\subsubsection{Stability of paths}
The conditions required for the existence and uniqueness in a time interval $[t_0, T]$ are the Lipschitz condition and the growth condition. Almost every stochastic differential equation encountered in practice satisfies the Lipschitz condition since it is essentially a smoothness condition. The growth condition is that there exists a $K$ such that for all $t$ in the range $[t_0, T]$
\begin{equation}
    |f(x,t)|^2 + |g(x,t)|^2 \leq K^2 (1 + |x|^2)
\end{equation}
The growth condition in contrast to the Lipschitz condition is often violated. If these conditions hold then it is guaranteed that the solution will not become infinite in a finite time. For our case of $n = 0$ and $m = 1$ it is guaranteed that the solution will not explode. This is however not guaranteed if either $n > 1$ or $m > 1$. Simulations suggest that the trajectory does not explode over not too time frames for $n$ or $m$ slightly greater than $1$. Nevertheless it is for this reason that we generally restrict ourselves to the cases $n \leq 1$ and $m \leq 1$.

\begin{figure*}
	\begin{center} 
		\hspace*{-0.5cm}\includegraphics[width=9cm] {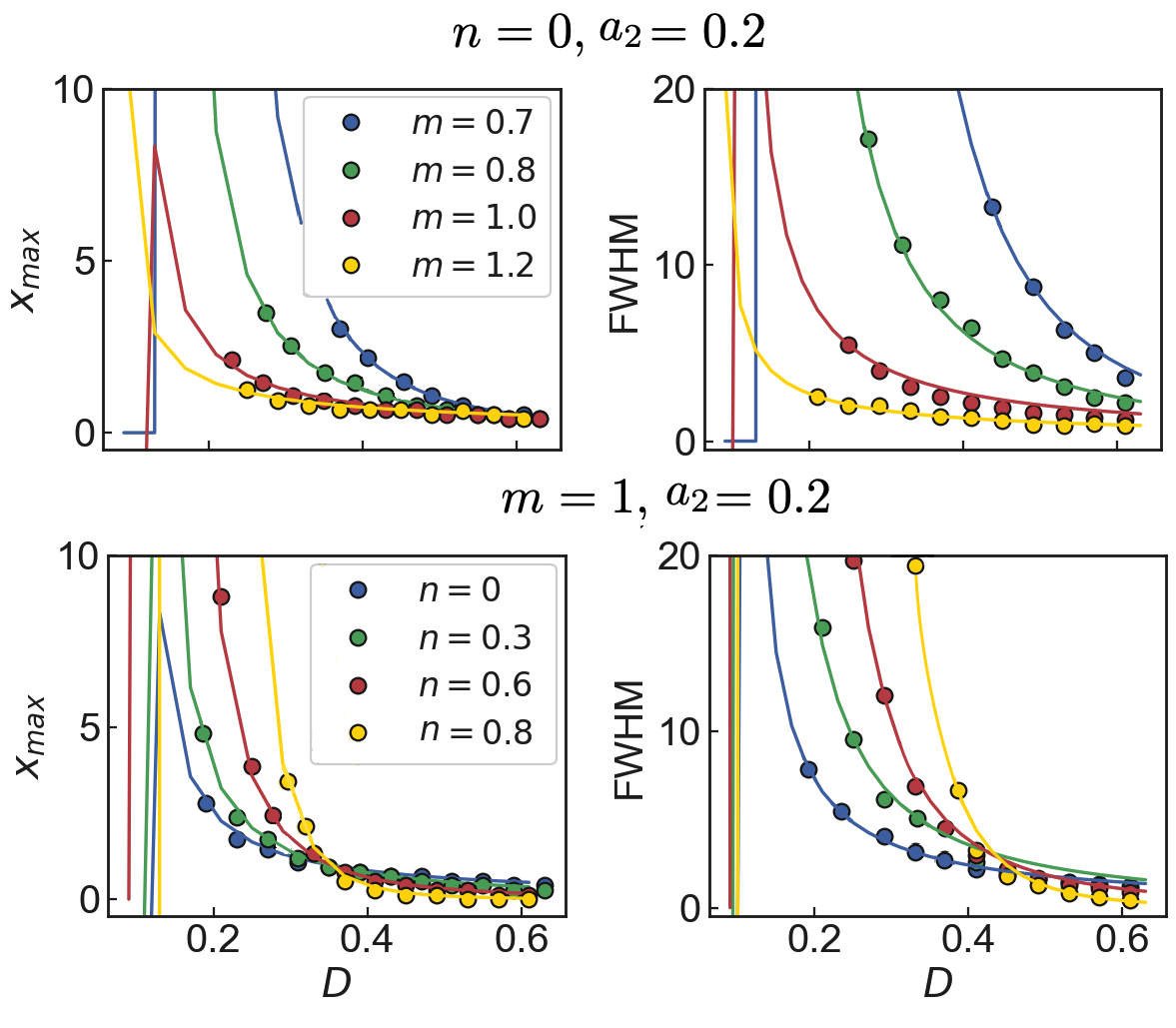}\caption{$x_{\text{max}}$ and FWHM of Eq. (\ref{sde w a2}) for $a_2 = 0.2$ and $a_1 = 0.5$. Lines of $x_{max}$ and FWHM are fits using Eqs. (\ref{max scale}) and (\ref{FWHM_gen}) respectively.}\label{fwhm_and_max_a2_gr0}
	\end{center}
\end{figure*}

\subsection{General}

\subsubsection{Results for $a_2 > 0$}
Results for $a_2 > 0$ are shown in Fig. \ref{fwhm_and_max_a2_gr0}. Agreement seems to exist between conjecture and simulations even for smaller $D$, although simulations become increasingly more challenge in this domain, requiring a longer simulation time and a smaller time step.

\subsubsection{Higher order 'M-moments'}
We also speculate that instead of higher order Gaussian moments such as the kurtosis, quantifying a heavy tailed distribution using the power law of the tail is more appropriate. The link between power laws and multiplicative noise has already been the work of recent literature \cite{levy1996power, sornette1997convergent, sornette1998multiplicative}. 

\bibliography{bibliography.bib}

\begin{thebibliography}{66}%
\makeatletter
\providecommand \@ifxundefined [1]{%
 \@ifx{#1\undefined}
}%
\providecommand \@ifnum [1]{%
 \ifnum #1\expandafter \@firstoftwo
 \else \expandafter \@secondoftwo
 \fi
}%
\providecommand \@ifx [1]{%
 \ifx #1\expandafter \@firstoftwo
 \else \expandafter \@secondoftwo
 \fi
}%
\providecommand \natexlab [1]{#1}%
\providecommand \enquote  [1]{``#1''}%
\providecommand \bibnamefont  [1]{#1}%
\providecommand \bibfnamefont [1]{#1}%
\providecommand \citenamefont [1]{#1}%
\providecommand \href@noop [0]{\@secondoftwo}%
\providecommand \href [0]{\begingroup \@sanitize@url \@href}%
\providecommand \@href[1]{\@@startlink{#1}\@@href}%
\providecommand \@@href[1]{\endgroup#1\@@endlink}%
\providecommand \@sanitize@url [0]{\catcode `\\12\catcode `\$12\catcode `\&12\catcode `\#12\catcode `\^12\catcode `\_12\catcode `\%12\relax}%
\providecommand \@@startlink[1]{}%
\providecommand \@@endlink[0]{}%
\providecommand \url  [0]{\begingroup\@sanitize@url \@url }%
\providecommand \@url [1]{\endgroup\@href {#1}{\urlprefix }}%
\providecommand \urlprefix  [0]{URL }%
\providecommand \Eprint [0]{\href }%
\providecommand \doibase [0]{https://doi.org/}%
\providecommand \selectlanguage [0]{\@gobble}%
\providecommand \bibinfo  [0]{\@secondoftwo}%
\providecommand \bibfield  [0]{\@secondoftwo}%
\providecommand \translation [1]{[#1]}%
\providecommand \BibitemOpen [0]{}%
\providecommand \bibitemStop [0]{}%
\providecommand \bibitemNoStop [0]{.\EOS\space}%
\providecommand \EOS [0]{\spacefactor3000\relax}%
\providecommand \BibitemShut  [1]{\csname bibitem#1\endcsname}%
\let\auto@bib@innerbib\@empty
\bibitem [{\citenamefont {Horsthemke}(1984)}]{horsthemke1984noise}%
  \BibitemOpen
  \bibfield  {author} {\bibinfo {author} {\bibfnamefont {W.}~\bibnamefont {Horsthemke}},\ }\bibfield  {title} {\bibinfo {title} {{Noise Induced Transitions}},\ }in\ \href@noop {} {\emph {\bibinfo {booktitle} {Non-Equilibrium Dynamics in Chemical Systems: Proceedings of the International Symposium, Bordeaux, France, September 3--7, 1984}}}\ (\bibinfo {organization} {Springer},\ \bibinfo {year} {1984})\ pp.\ \bibinfo {pages} {150--160}\BibitemShut {NoStop}%
\bibitem [{\citenamefont {Feudel}\ and\ \citenamefont {Grebogi}(1997)}]{feudel1997multistability}%
  \BibitemOpen
  \bibfield  {author} {\bibinfo {author} {\bibfnamefont {U.}~\bibnamefont {Feudel}}\ and\ \bibinfo {author} {\bibfnamefont {C.}~\bibnamefont {Grebogi}},\ }\bibfield  {title} {\bibinfo {title} {{Multistability and the Control of Complexity}},\ }\href@noop {} {\bibfield  {journal} {\bibinfo  {journal} {Chaos}\ }\textbf {\bibinfo {volume} {7}},\ \bibinfo {pages} {597} (\bibinfo {year} {1997})}\BibitemShut {NoStop}%
\bibitem [{\citenamefont {Pisarchik}\ and\ \citenamefont {Feudel}(2014)}]{pisarchik2014control}%
  \BibitemOpen
  \bibfield  {author} {\bibinfo {author} {\bibfnamefont {A.~N.}\ \bibnamefont {Pisarchik}}\ and\ \bibinfo {author} {\bibfnamefont {U.}~\bibnamefont {Feudel}},\ }\bibfield  {title} {\bibinfo {title} {{Control of Multistability}},\ }\href@noop {} {\bibfield  {journal} {\bibinfo  {journal} {Physics Reports}\ }\textbf {\bibinfo {volume} {540}},\ \bibinfo {pages} {167} (\bibinfo {year} {2014})}\BibitemShut {NoStop}%
\bibitem [{\citenamefont {Feudel}(2008)}]{feudel2008complex}%
  \BibitemOpen
  \bibfield  {author} {\bibinfo {author} {\bibfnamefont {U.}~\bibnamefont {Feudel}},\ }\bibfield  {title} {\bibinfo {title} {{Complex Dynamics in Multistable Systems}},\ }\href@noop {} {\bibfield  {journal} {\bibinfo  {journal} {Int. J. Bifurcation Chaos}\ }\textbf {\bibinfo {volume} {18}},\ \bibinfo {pages} {1607} (\bibinfo {year} {2008})}\BibitemShut {NoStop}%
\bibitem [{\citenamefont {Kraut}\ and\ \citenamefont {Feudel}(2002)}]{kraut2002multistability}%
  \BibitemOpen
  \bibfield  {author} {\bibinfo {author} {\bibfnamefont {S.}~\bibnamefont {Kraut}}\ and\ \bibinfo {author} {\bibfnamefont {U.}~\bibnamefont {Feudel}},\ }\bibfield  {title} {\bibinfo {title} {{Multistability, Noise, and Attractor Hopping: The Crucial Role of Chaotic Saddles}},\ }\href@noop {} {\bibfield  {journal} {\bibinfo  {journal} {Phys. Rev. E}\ }\textbf {\bibinfo {volume} {66}},\ \bibinfo {pages} {015207} (\bibinfo {year} {2002})}\BibitemShut {NoStop}%
\bibitem [{\citenamefont {Kraut}\ \emph {et~al.}(1999)\citenamefont {Kraut}, \citenamefont {Feudel},\ and\ \citenamefont {Grebogi}}]{kraut1999preference}%
  \BibitemOpen
  \bibfield  {author} {\bibinfo {author} {\bibfnamefont {S.}~\bibnamefont {Kraut}}, \bibinfo {author} {\bibfnamefont {U.}~\bibnamefont {Feudel}},\ and\ \bibinfo {author} {\bibfnamefont {C.}~\bibnamefont {Grebogi}},\ }\bibfield  {title} {\bibinfo {title} {{Preference of Attractors in Noisy Multistable Systems}},\ }\href@noop {} {\bibfield  {journal} {\bibinfo  {journal} {Phys. Rev. E}\ }\textbf {\bibinfo {volume} {59}},\ \bibinfo {pages} {5253} (\bibinfo {year} {1999})}\BibitemShut {NoStop}%
\bibitem [{\citenamefont {Huerta-Cuellar}\ \emph {et~al.}(2008)\citenamefont {Huerta-Cuellar}, \citenamefont {Pisarchik},\ and\ \citenamefont {Barmenkov}}]{huerta2008experimental}%
  \BibitemOpen
  \bibfield  {author} {\bibinfo {author} {\bibfnamefont {G.}~\bibnamefont {Huerta-Cuellar}}, \bibinfo {author} {\bibfnamefont {A.~N.}\ \bibnamefont {Pisarchik}},\ and\ \bibinfo {author} {\bibfnamefont {Y.~O.}\ \bibnamefont {Barmenkov}},\ }\bibfield  {title} {\bibinfo {title} {{Experimental Characterization of Hopping Dynamics in a Multistable Fiber Laser}},\ }\href@noop {} {\bibfield  {journal} {\bibinfo  {journal} {Phys. Rev. E}\ }\textbf {\bibinfo {volume} {78}},\ \bibinfo {pages} {035202} (\bibinfo {year} {2008})}\BibitemShut {NoStop}%
\bibitem [{\citenamefont {de~Souza}\ \emph {et~al.}(2007)\citenamefont {de~Souza}, \citenamefont {Batista}, \citenamefont {Caldas}, \citenamefont {Viana},\ and\ \citenamefont {Kapitaniak}}]{de2007noise}%
  \BibitemOpen
  \bibfield  {author} {\bibinfo {author} {\bibfnamefont {S.~L.}\ \bibnamefont {de~Souza}}, \bibinfo {author} {\bibfnamefont {A.~M.}\ \bibnamefont {Batista}}, \bibinfo {author} {\bibfnamefont {I.~L.}\ \bibnamefont {Caldas}}, \bibinfo {author} {\bibfnamefont {R.~L.}\ \bibnamefont {Viana}},\ and\ \bibinfo {author} {\bibfnamefont {T.}~\bibnamefont {Kapitaniak}},\ }\bibfield  {title} {\bibinfo {title} {{Noise-Induced Basin Hopping in a Vibro-Impact System}},\ }\href@noop {} {\bibfield  {journal} {\bibinfo  {journal} {Chaos, Solitons \& Fractals}\ }\textbf {\bibinfo {volume} {32}},\ \bibinfo {pages} {758} (\bibinfo {year} {2007})}\BibitemShut {NoStop}%
\bibitem [{\citenamefont {Landa}\ and\ \citenamefont {McClintock}(2000)}]{landa2000changes}%
  \BibitemOpen
  \bibfield  {author} {\bibinfo {author} {\bibfnamefont {P.~S.}\ \bibnamefont {Landa}}\ and\ \bibinfo {author} {\bibfnamefont {P.~V.~E.}\ \bibnamefont {McClintock}},\ }\bibfield  {title} {\bibinfo {title} {{Changes in the Dynamical Behavior of Nonlinear Systems Induced by Noise}},\ }\href@noop {} {\bibfield  {journal} {\bibinfo  {journal} {Physics Reports}\ }\textbf {\bibinfo {volume} {323}},\ \bibinfo {pages} {1} (\bibinfo {year} {2000})}\BibitemShut {NoStop}%
\bibitem [{\citenamefont {Forgoston}\ and\ \citenamefont {Moore}(2018)}]{forgoston2018primer}%
  \BibitemOpen
  \bibfield  {author} {\bibinfo {author} {\bibfnamefont {E.}~\bibnamefont {Forgoston}}\ and\ \bibinfo {author} {\bibfnamefont {R.~O.}\ \bibnamefont {Moore}},\ }\bibfield  {title} {\bibinfo {title} {{A Primer on Noise-Induced Transitions in Applied Dynamical Systems}},\ }\href@noop {} {\bibfield  {journal} {\bibinfo  {journal} {SIAM Review}\ }\textbf {\bibinfo {volume} {60}},\ \bibinfo {pages} {969} (\bibinfo {year} {2018})}\BibitemShut {NoStop}%
\bibitem [{\citenamefont {Van~den Broeck}\ \emph {et~al.}(1994)\citenamefont {Van~den Broeck}, \citenamefont {Parrondo},\ and\ \citenamefont {Toral}}]{van1994noise}%
  \BibitemOpen
  \bibfield  {author} {\bibinfo {author} {\bibfnamefont {C.}~\bibnamefont {Van~den Broeck}}, \bibinfo {author} {\bibfnamefont {J.}~\bibnamefont {Parrondo}},\ and\ \bibinfo {author} {\bibfnamefont {R.}~\bibnamefont {Toral}},\ }\bibfield  {title} {\bibinfo {title} {{Noise-Induced Nonequilibrium Phase Transition}},\ }\href@noop {} {\bibfield  {journal} {\bibinfo  {journal} {Phys. Rev. Lett.}\ }\textbf {\bibinfo {volume} {73}},\ \bibinfo {pages} {3395} (\bibinfo {year} {1994})}\BibitemShut {NoStop}%
\bibitem [{\citenamefont {Hodgkinson}\ and\ \citenamefont {Mahoney}(2021)}]{hodgkinson2021multiplicative}%
  \BibitemOpen
  \bibfield  {author} {\bibinfo {author} {\bibfnamefont {L.}~\bibnamefont {Hodgkinson}}\ and\ \bibinfo {author} {\bibfnamefont {M.}~\bibnamefont {Mahoney}},\ }\bibfield  {title} {\bibinfo {title} {{Multiplicative Noise and Heavy Tails in Stochastic Optimization}},\ }in\ \href@noop {} {\emph {\bibinfo {booktitle} {International Conference on Machine Learning}}}\ (\bibinfo {organization} {PMLR},\ \bibinfo {year} {2021})\ pp.\ \bibinfo {pages} {4262--4274}\BibitemShut {NoStop}%
\bibitem [{\citenamefont {Kim}\ \emph {et~al.}(1997{\natexlab{a}})\citenamefont {Kim}, \citenamefont {Park},\ and\ \citenamefont {Ryu}}]{kim1997noise}%
  \BibitemOpen
  \bibfield  {author} {\bibinfo {author} {\bibfnamefont {S.}~\bibnamefont {Kim}}, \bibinfo {author} {\bibfnamefont {S.~H.}\ \bibnamefont {Park}},\ and\ \bibinfo {author} {\bibfnamefont {C.~S.}\ \bibnamefont {Ryu}},\ }\bibfield  {title} {\bibinfo {title} {{Noise-Enhanced Multistability in Coupled Oscillator Systems}},\ }\href@noop {} {\bibfield  {journal} {\bibinfo  {journal} {Phys. Rev. Lett.}\ }\textbf {\bibinfo {volume} {78}},\ \bibinfo {pages} {1616} (\bibinfo {year} {1997}{\natexlab{a}})}\BibitemShut {NoStop}%
\bibitem [{\citenamefont {Kim}\ \emph {et~al.}(1997{\natexlab{b}})\citenamefont {Kim}, \citenamefont {Park}, \citenamefont {Doering},\ and\ \citenamefont {Ryu}}]{kim1997reentrant}%
  \BibitemOpen
  \bibfield  {author} {\bibinfo {author} {\bibfnamefont {S.}~\bibnamefont {Kim}}, \bibinfo {author} {\bibfnamefont {S.~H.}\ \bibnamefont {Park}}, \bibinfo {author} {\bibfnamefont {C.~R.}\ \bibnamefont {Doering}},\ and\ \bibinfo {author} {\bibfnamefont {C.~S.}\ \bibnamefont {Ryu}},\ }\bibfield  {title} {\bibinfo {title} {{Reentrant Transitions in Globally Coupled Active Rotators With Multiplicative and Additive Noises}},\ }\href@noop {} {\bibfield  {journal} {\bibinfo  {journal} {Phys. Lett. A}\ }\textbf {\bibinfo {volume} {224}},\ \bibinfo {pages} {147} (\bibinfo {year} {1997}{\natexlab{b}})}\BibitemShut {NoStop}%
\bibitem [{\citenamefont {Lee}\ \emph {et~al.}(1998)\citenamefont {Lee}, \citenamefont {Kwak},\ and\ \citenamefont {Lim}}]{lee1998phase}%
  \BibitemOpen
  \bibfield  {author} {\bibinfo {author} {\bibfnamefont {K.~J.}\ \bibnamefont {Lee}}, \bibinfo {author} {\bibfnamefont {Y.}~\bibnamefont {Kwak}},\ and\ \bibinfo {author} {\bibfnamefont {T.~K.}\ \bibnamefont {Lim}},\ }\bibfield  {title} {\bibinfo {title} {{Phase Jumps Near a Phase Synchronization Transition in Systems of Two Coupled Chaotic Oscillators}},\ }\href@noop {} {\bibfield  {journal} {\bibinfo  {journal} {Phys. Rev. Lett.}\ }\textbf {\bibinfo {volume} {81}},\ \bibinfo {pages} {321} (\bibinfo {year} {1998})}\BibitemShut {NoStop}%
\bibitem [{\citenamefont {Boccaletti}\ \emph {et~al.}(2002)\citenamefont {Boccaletti}, \citenamefont {Kurths}, \citenamefont {Osipov}, \citenamefont {Valladares},\ and\ \citenamefont {Zhou}}]{boccaletti2002synchronization}%
  \BibitemOpen
  \bibfield  {author} {\bibinfo {author} {\bibfnamefont {S.}~\bibnamefont {Boccaletti}}, \bibinfo {author} {\bibfnamefont {J.}~\bibnamefont {Kurths}}, \bibinfo {author} {\bibfnamefont {G.}~\bibnamefont {Osipov}}, \bibinfo {author} {\bibfnamefont {D.}~\bibnamefont {Valladares}},\ and\ \bibinfo {author} {\bibfnamefont {C.}~\bibnamefont {Zhou}},\ }\bibfield  {title} {\bibinfo {title} {{The Synchronization of Chaotic Systems}},\ }\href@noop {} {\bibfield  {journal} {\bibinfo  {journal} {Physics reports}\ }\textbf {\bibinfo {volume} {366}},\ \bibinfo {pages} {1} (\bibinfo {year} {2002})}\BibitemShut {NoStop}%
\bibitem [{\citenamefont {Cabrera}\ and\ \citenamefont {Milton}(2002)}]{cabrera2002off}%
  \BibitemOpen
  \bibfield  {author} {\bibinfo {author} {\bibfnamefont {J.~L.}\ \bibnamefont {Cabrera}}\ and\ \bibinfo {author} {\bibfnamefont {J.~G.}\ \bibnamefont {Milton}},\ }\bibfield  {title} {\bibinfo {title} {{On-Off Intermittency in a Human Balancing Task}},\ }\href@noop {} {\bibfield  {journal} {\bibinfo  {journal} {Phys. Rev. Lett.}\ }\textbf {\bibinfo {volume} {89}},\ \bibinfo {pages} {158702} (\bibinfo {year} {2002})}\BibitemShut {NoStop}%
\bibitem [{\citenamefont {Krawiecki}\ \emph {et~al.}(2002)\citenamefont {Krawiecki}, \citenamefont {Ho{\l}yst},\ and\ \citenamefont {Helbing}}]{krawiecki2002volatility}%
  \BibitemOpen
  \bibfield  {author} {\bibinfo {author} {\bibfnamefont {A.}~\bibnamefont {Krawiecki}}, \bibinfo {author} {\bibfnamefont {J.}~\bibnamefont {Ho{\l}yst}},\ and\ \bibinfo {author} {\bibfnamefont {D.}~\bibnamefont {Helbing}},\ }\bibfield  {title} {\bibinfo {title} {{Volatility Clustering and Scaling for Financial Time Series Due to Attractor Bubbling}},\ }\href@noop {} {\bibfield  {journal} {\bibinfo  {journal} {Phys. Rev. Lett.}\ }\textbf {\bibinfo {volume} {89}},\ \bibinfo {pages} {158701} (\bibinfo {year} {2002})}\BibitemShut {NoStop}%
\bibitem [{\citenamefont {Mandelbrot}\ and\ \citenamefont {Mandelbrot}(1997)}]{mandelbrot1997variation}%
  \BibitemOpen
  \bibfield  {author} {\bibinfo {author} {\bibfnamefont {B.~B.}\ \bibnamefont {Mandelbrot}}\ and\ \bibinfo {author} {\bibfnamefont {B.~B.}\ \bibnamefont {Mandelbrot}},\ }\href@noop {} {\emph {\bibinfo {title} {{The Variation of Certain Speculative Prices}}}}\ (\bibinfo  {publisher} {Springer},\ \bibinfo {year} {1997})\BibitemShut {NoStop}%
\bibitem [{\citenamefont {Imkeller}\ and\ \citenamefont {Von~Storch}(2001)}]{imkeller2001stochastic}%
  \BibitemOpen
  \bibfield  {author} {\bibinfo {author} {\bibfnamefont {P.}~\bibnamefont {Imkeller}}\ and\ \bibinfo {author} {\bibfnamefont {J.-S.}\ \bibnamefont {Von~Storch}},\ }\href@noop {} {\emph {\bibinfo {title} {{Stochastic Climate Models}}}},\ Vol.~\bibinfo {volume} {49}\ (\bibinfo  {publisher} {Springer Science \& Business Media},\ \bibinfo {year} {2001})\BibitemShut {NoStop}%
\bibitem [{\citenamefont {IPCC}(2021)}]{RN4}%
  \BibitemOpen
  \bibfield  {author} {\bibinfo {author} {\bibnamefont {IPCC}},\ }\bibinfo {title} {Technical summary},\ in\ \href {https://doi.org/10.1017/9781009157896.002} {\emph {\bibinfo {booktitle} {{Climate Change 2021: The Physical Science Basis. Contribution of Working Group I to the Sixth Assessment Report of the Intergovernmental Panel on Climate Change}}}},\ \bibinfo {editor} {edited by\ \bibinfo {editor} {\bibfnamefont {V.}~\bibnamefont {Masson-Delmotte}}, \bibinfo {editor} {\bibfnamefont {P.}~\bibnamefont {Zhai}}, \bibinfo {editor} {\bibfnamefont {A.}~\bibnamefont {Pirani}}, \bibinfo {editor} {\bibfnamefont {S.}~\bibnamefont {Connors}}, \bibinfo {editor} {\bibfnamefont {C.}~\bibnamefont {Péan}}, \bibinfo {editor} {\bibfnamefont {S.}~\bibnamefont {Berger}}, \bibinfo {editor} {\bibfnamefont {N.}~\bibnamefont {Caud}}, \bibinfo {editor} {\bibfnamefont {Y.}~\bibnamefont {Chen}}, \bibinfo {editor} {\bibfnamefont {L.}~\bibnamefont {Goldfarb}}, \bibinfo {editor} {\bibfnamefont {M.}~\bibnamefont {Gomis}}, \bibinfo
  {editor} {\bibfnamefont {M.}~\bibnamefont {Huang}}, \bibinfo {editor} {\bibfnamefont {K.}~\bibnamefont {Leitzell}}, \bibinfo {editor} {\bibfnamefont {E.}~\bibnamefont {Lonnoy}}, \bibinfo {editor} {\bibfnamefont {J.}~\bibnamefont {Matthews}}, \bibinfo {editor} {\bibfnamefont {T.}~\bibnamefont {Maycock}}, \bibinfo {editor} {\bibfnamefont {T.}~\bibnamefont {Waterfield}}, \bibinfo {editor} {\bibfnamefont {O.}~\bibnamefont {Yelekçi}}, \bibinfo {editor} {\bibfnamefont {R.}~\bibnamefont {Yu}},\ and\ \bibinfo {editor} {\bibfnamefont {B.}~\bibnamefont {Zhou}}}\ (\bibinfo  {publisher} {Cambridge University Press},\ \bibinfo {address} {Cambridge, United Kingdom and New York, NY, USA},\ \bibinfo {year} {2021})\ p.\ \bibinfo {pages} {33−144}\BibitemShut {NoStop}%
\bibitem [{\citenamefont {Majda}\ \emph {et~al.}(1999)\citenamefont {Majda}, \citenamefont {Timofeyev},\ and\ \citenamefont {Vanden~Eijnden}}]{majda1999models}%
  \BibitemOpen
  \bibfield  {author} {\bibinfo {author} {\bibfnamefont {A.~J.}\ \bibnamefont {Majda}}, \bibinfo {author} {\bibfnamefont {I.}~\bibnamefont {Timofeyev}},\ and\ \bibinfo {author} {\bibfnamefont {E.}~\bibnamefont {Vanden~Eijnden}},\ }\bibfield  {title} {\bibinfo {title} {{Models for Stochastic Climate Prediction}},\ }\href@noop {} {\bibfield  {journal} {\bibinfo  {journal} {PNAS}\ }\textbf {\bibinfo {volume} {96}},\ \bibinfo {pages} {14687} (\bibinfo {year} {1999})}\BibitemShut {NoStop}%
\bibitem [{\citenamefont {Majda}\ \emph {et~al.}(2009)\citenamefont {Majda}, \citenamefont {Franzke},\ and\ \citenamefont {Crommelin}}]{majda2009normal}%
  \BibitemOpen
  \bibfield  {author} {\bibinfo {author} {\bibfnamefont {A.~J.}\ \bibnamefont {Majda}}, \bibinfo {author} {\bibfnamefont {C.}~\bibnamefont {Franzke}},\ and\ \bibinfo {author} {\bibfnamefont {D.}~\bibnamefont {Crommelin}},\ }\bibfield  {title} {\bibinfo {title} {{Normal Forms for Reduced Stochastic Climate Models}},\ }\href@noop {} {\bibfield  {journal} {\bibinfo  {journal} {PNAS}\ }\textbf {\bibinfo {volume} {106}},\ \bibinfo {pages} {3649} (\bibinfo {year} {2009})}\BibitemShut {NoStop}%
\bibitem [{\citenamefont {Sardeshmukh}\ and\ \citenamefont {Sura}(2009)}]{sardeshmukh2009reconciling}%
  \BibitemOpen
  \bibfield  {author} {\bibinfo {author} {\bibfnamefont {P.~D.}\ \bibnamefont {Sardeshmukh}}\ and\ \bibinfo {author} {\bibfnamefont {P.}~\bibnamefont {Sura}},\ }\bibfield  {title} {\bibinfo {title} {{Reconciling Non Gaussian Climate Statistics with Linear Dynamics}},\ }\href@noop {} {\bibfield  {journal} {\bibinfo  {journal} {J. Clim.}\ }\textbf {\bibinfo {volume} {22}},\ \bibinfo {pages} {1193} (\bibinfo {year} {2009})}\BibitemShut {NoStop}%
\bibitem [{\citenamefont {Berner}\ \emph {et~al.}(2017)\citenamefont {Berner}, \citenamefont {Achatz}, \citenamefont {Batte}, \citenamefont {Bengtsson}, \citenamefont {C{\'a}mara}, \citenamefont {Christensen}, \citenamefont {Colangeli}, \citenamefont {Coleman}, \citenamefont {Crommelin}, \citenamefont {Dolaptchiev} \emph {et~al.}}]{berner2017stochastic}%
  \BibitemOpen
  \bibfield  {author} {\bibinfo {author} {\bibfnamefont {J.}~\bibnamefont {Berner}}, \bibinfo {author} {\bibfnamefont {U.}~\bibnamefont {Achatz}}, \bibinfo {author} {\bibfnamefont {L.}~\bibnamefont {Batte}}, \bibinfo {author} {\bibfnamefont {L.}~\bibnamefont {Bengtsson}}, \bibinfo {author} {\bibfnamefont {A.~d.~l.}\ \bibnamefont {C{\'a}mara}}, \bibinfo {author} {\bibfnamefont {H.~M.}\ \bibnamefont {Christensen}}, \bibinfo {author} {\bibfnamefont {M.}~\bibnamefont {Colangeli}}, \bibinfo {author} {\bibfnamefont {D.~R.}\ \bibnamefont {Coleman}}, \bibinfo {author} {\bibfnamefont {D.}~\bibnamefont {Crommelin}}, \bibinfo {author} {\bibfnamefont {S.~I.}\ \bibnamefont {Dolaptchiev}}, \emph {et~al.},\ }\bibfield  {title} {\bibinfo {title} {{Stochastic Parameterization: Toward a New View of Weather and Climate Models}},\ }\href@noop {} {\bibfield  {journal} {\bibinfo  {journal} {Bull. Am. Meteorol. Soc.}\ }\textbf {\bibinfo {volume} {98}},\ \bibinfo {pages} {565} (\bibinfo {year} {2017})}\BibitemShut {NoStop}%
\bibitem [{\citenamefont {Sura}\ \emph {et~al.}(2005)\citenamefont {Sura}, \citenamefont {Newman}, \citenamefont {Penland},\ and\ \citenamefont {Sardeshmukh}}]{sura2005multiplicative}%
  \BibitemOpen
  \bibfield  {author} {\bibinfo {author} {\bibfnamefont {P.}~\bibnamefont {Sura}}, \bibinfo {author} {\bibfnamefont {M.}~\bibnamefont {Newman}}, \bibinfo {author} {\bibfnamefont {C.}~\bibnamefont {Penland}},\ and\ \bibinfo {author} {\bibfnamefont {P.}~\bibnamefont {Sardeshmukh}},\ }\bibfield  {title} {\bibinfo {title} {{Multiplicative Noise and Non-Gaussianity: A Paradigm for Atmospheric Regimes?}},\ }\href@noop {} {\bibfield  {journal} {\bibinfo  {journal} {J. Atmos. Sci.}\ }\textbf {\bibinfo {volume} {62}},\ \bibinfo {pages} {1391} (\bibinfo {year} {2005})}\BibitemShut {NoStop}%
\bibitem [{\citenamefont {Buizza}\ \emph {et~al.}(1999)\citenamefont {Buizza}, \citenamefont {Milleer},\ and\ \citenamefont {Palmer}}]{buizza1999stochastic}%
  \BibitemOpen
  \bibfield  {author} {\bibinfo {author} {\bibfnamefont {R.}~\bibnamefont {Buizza}}, \bibinfo {author} {\bibfnamefont {M.}~\bibnamefont {Milleer}},\ and\ \bibinfo {author} {\bibfnamefont {T.~N.}\ \bibnamefont {Palmer}},\ }\bibfield  {title} {\bibinfo {title} {{Stochastic Representation of Model Uncertainties in the ECMWF Ensemble Prediction System}},\ }\href@noop {} {\bibfield  {journal} {\bibinfo  {journal} {Q. J. R. Meteorolog. Soc.}\ }\textbf {\bibinfo {volume} {125}},\ \bibinfo {pages} {2887} (\bibinfo {year} {1999})}\BibitemShut {NoStop}%
\bibitem [{\citenamefont {Wang}\ and\ \citenamefont {Aljadeff}(2022)}]{wang2022multiplicative}%
  \BibitemOpen
  \bibfield  {author} {\bibinfo {author} {\bibfnamefont {B.}~\bibnamefont {Wang}}\ and\ \bibinfo {author} {\bibfnamefont {J.}~\bibnamefont {Aljadeff}},\ }\bibfield  {title} {\bibinfo {title} {{Multiplicative Shot-Noise: A New Route to Stability of Plastic Networks}},\ }\href@noop {} {\bibfield  {journal} {\bibinfo  {journal} {Phys. Rev. Lett.}\ }\textbf {\bibinfo {volume} {129}},\ \bibinfo {pages} {068101} (\bibinfo {year} {2022})}\BibitemShut {NoStop}%
\bibitem [{\citenamefont {Bauermann}\ and\ \citenamefont {Lindner}(2019)}]{bauermann2019multiplicative}%
  \BibitemOpen
  \bibfield  {author} {\bibinfo {author} {\bibfnamefont {J.}~\bibnamefont {Bauermann}}\ and\ \bibinfo {author} {\bibfnamefont {B.}~\bibnamefont {Lindner}},\ }\bibfield  {title} {\bibinfo {title} {{Multiplicative Noise Is Beneficial for the Transmission of Sensory Signals in Simple Neuron Models}},\ }\href@noop {} {\bibfield  {journal} {\bibinfo  {journal} {Biosystems}\ }\textbf {\bibinfo {volume} {178}},\ \bibinfo {pages} {25} (\bibinfo {year} {2019})}\BibitemShut {NoStop}%
\bibitem [{\citenamefont {Pavlyukevich}(2007)}]{pavlyukevich2007levy}%
  \BibitemOpen
  \bibfield  {author} {\bibinfo {author} {\bibfnamefont {I.}~\bibnamefont {Pavlyukevich}},\ }\bibfield  {title} {\bibinfo {title} {{L{\'e}vy Flights, Non-Local Search and Simulated Annealing}},\ }\href@noop {} {\bibfield  {journal} {\bibinfo  {journal} {journal of computational physics}\ }\textbf {\bibinfo {volume} {226}},\ \bibinfo {pages} {1830} (\bibinfo {year} {2007})}\BibitemShut {NoStop}%
\bibitem [{\citenamefont {Platt}\ \emph {et~al.}(1993)\citenamefont {Platt}, \citenamefont {Spiegel},\ and\ \citenamefont {Tresser}}]{platt1993off}%
  \BibitemOpen
  \bibfield  {author} {\bibinfo {author} {\bibfnamefont {N.}~\bibnamefont {Platt}}, \bibinfo {author} {\bibfnamefont {E.}~\bibnamefont {Spiegel}},\ and\ \bibinfo {author} {\bibfnamefont {C.}~\bibnamefont {Tresser}},\ }\bibfield  {title} {\bibinfo {title} {{On-off Intermittency: A Mechanism for Bursting}},\ }\href@noop {} {\bibfield  {journal} {\bibinfo  {journal} {Phys. Rev. Lett.}\ }\textbf {\bibinfo {volume} {70}},\ \bibinfo {pages} {279} (\bibinfo {year} {1993})}\BibitemShut {NoStop}%
\bibitem [{\citenamefont {Heagy}\ \emph {et~al.}(1994)\citenamefont {Heagy}, \citenamefont {Platt},\ and\ \citenamefont {Hammel}}]{heagy1994characterization}%
  \BibitemOpen
  \bibfield  {author} {\bibinfo {author} {\bibfnamefont {J.}~\bibnamefont {Heagy}}, \bibinfo {author} {\bibfnamefont {N.}~\bibnamefont {Platt}},\ and\ \bibinfo {author} {\bibfnamefont {S.}~\bibnamefont {Hammel}},\ }\bibfield  {title} {\bibinfo {title} {{Characterization of On-Off Intermittency}},\ }\href@noop {} {\bibfield  {journal} {\bibinfo  {journal} {Phys. Rev. E}\ }\textbf {\bibinfo {volume} {49}},\ \bibinfo {pages} {1140} (\bibinfo {year} {1994})}\BibitemShut {NoStop}%
\bibitem [{\citenamefont {Ding}\ and\ \citenamefont {Yang}(1997)}]{ding1997stability}%
  \BibitemOpen
  \bibfield  {author} {\bibinfo {author} {\bibfnamefont {M.}~\bibnamefont {Ding}}\ and\ \bibinfo {author} {\bibfnamefont {W.}~\bibnamefont {Yang}},\ }\bibfield  {title} {\bibinfo {title} {{Stability of Synchronous Chaos and On-Off Intermittency in Coupled Map Lattices}},\ }\href@noop {} {\bibfield  {journal} {\bibinfo  {journal} {Phys. Rev. E}\ }\textbf {\bibinfo {volume} {56}},\ \bibinfo {pages} {4009} (\bibinfo {year} {1997})}\BibitemShut {NoStop}%
\bibitem [{\citenamefont {Elaskar}\ and\ \citenamefont {del R{\'\i}o}(2023)}]{elaskar2023review}%
  \BibitemOpen
  \bibfield  {author} {\bibinfo {author} {\bibfnamefont {S.}~\bibnamefont {Elaskar}}\ and\ \bibinfo {author} {\bibfnamefont {E.}~\bibnamefont {del R{\'\i}o}},\ }\bibfield  {title} {\bibinfo {title} {{Review of Chaotic Intermittency}},\ }\href@noop {} {\bibfield  {journal} {\bibinfo  {journal} {Symmetry}\ }\textbf {\bibinfo {volume} {15}},\ \bibinfo {pages} {1195} (\bibinfo {year} {2023})}\BibitemShut {NoStop}%
\bibitem [{\citenamefont {Ashwin}\ \emph {et~al.}(1994)\citenamefont {Ashwin}, \citenamefont {Buescu},\ and\ \citenamefont {Stewart}}]{ashwin1994bubbling}%
  \BibitemOpen
  \bibfield  {author} {\bibinfo {author} {\bibfnamefont {P.}~\bibnamefont {Ashwin}}, \bibinfo {author} {\bibfnamefont {J.}~\bibnamefont {Buescu}},\ and\ \bibinfo {author} {\bibfnamefont {I.}~\bibnamefont {Stewart}},\ }\bibfield  {title} {\bibinfo {title} {{Bubbling of Attractors and Synchronisation of Chaotic Oscillators}},\ }\href@noop {} {\bibfield  {journal} {\bibinfo  {journal} {Phys. Lett. A}\ }\textbf {\bibinfo {volume} {193}},\ \bibinfo {pages} {126} (\bibinfo {year} {1994})}\BibitemShut {NoStop}%
\bibitem [{\citenamefont {Hramov}\ \emph {et~al.}(2006)\citenamefont {Hramov}, \citenamefont {Koronovskii}, \citenamefont {Kurovskaya},\ and\ \citenamefont {Boccaletti}}]{hramov2006ring}%
  \BibitemOpen
  \bibfield  {author} {\bibinfo {author} {\bibfnamefont {A.~E.}\ \bibnamefont {Hramov}}, \bibinfo {author} {\bibfnamefont {A.~A.}\ \bibnamefont {Koronovskii}}, \bibinfo {author} {\bibfnamefont {M.~K.}\ \bibnamefont {Kurovskaya}},\ and\ \bibinfo {author} {\bibfnamefont {S.}~\bibnamefont {Boccaletti}},\ }\bibfield  {title} {\bibinfo {title} {{Ring Intermittency in Coupled Chaotic Oscillators at the Boundary of Phase Synchronization}},\ }\href@noop {} {\bibfield  {journal} {\bibinfo  {journal} {Phys. Rev. Lett.}\ }\textbf {\bibinfo {volume} {97}},\ \bibinfo {pages} {114101} (\bibinfo {year} {2006})}\BibitemShut {NoStop}%
\bibitem [{\citenamefont {Scheffer}\ \emph {et~al.}(2009)\citenamefont {Scheffer}, \citenamefont {Bascompte}, \citenamefont {Brock}, \citenamefont {Brovkin}, \citenamefont {Carpenter}, \citenamefont {Dakos}, \citenamefont {Held}, \citenamefont {Van~Nes}, \citenamefont {Rietkerk},\ and\ \citenamefont {Sugihara}}]{scheffer2009early}%
  \BibitemOpen
  \bibfield  {author} {\bibinfo {author} {\bibfnamefont {M.}~\bibnamefont {Scheffer}}, \bibinfo {author} {\bibfnamefont {J.}~\bibnamefont {Bascompte}}, \bibinfo {author} {\bibfnamefont {W.~A.}\ \bibnamefont {Brock}}, \bibinfo {author} {\bibfnamefont {V.}~\bibnamefont {Brovkin}}, \bibinfo {author} {\bibfnamefont {S.~R.}\ \bibnamefont {Carpenter}}, \bibinfo {author} {\bibfnamefont {V.}~\bibnamefont {Dakos}}, \bibinfo {author} {\bibfnamefont {H.}~\bibnamefont {Held}}, \bibinfo {author} {\bibfnamefont {E.~H.}\ \bibnamefont {Van~Nes}}, \bibinfo {author} {\bibfnamefont {M.}~\bibnamefont {Rietkerk}},\ and\ \bibinfo {author} {\bibfnamefont {G.}~\bibnamefont {Sugihara}},\ }\bibfield  {title} {\bibinfo {title} {{Early-Warning Signals for Critical Transitions}},\ }\href@noop {} {\bibfield  {journal} {\bibinfo  {journal} {Nature}\ }\textbf {\bibinfo {volume} {461}},\ \bibinfo {pages} {53} (\bibinfo {year} {2009})}\BibitemShut {NoStop}%
\bibitem [{\citenamefont {Young}\ and\ \citenamefont {Singh}(1988)}]{young1988statistical}%
  \BibitemOpen
  \bibfield  {author} {\bibinfo {author} {\bibfnamefont {M.~R.}\ \bibnamefont {Young}}\ and\ \bibinfo {author} {\bibfnamefont {S.}~\bibnamefont {Singh}},\ }\bibfield  {title} {\bibinfo {title} {{Statistical Properties of a Laser With Multiplicative Noise}},\ }\href@noop {} {\bibfield  {journal} {\bibinfo  {journal} {Opt. lett.}\ }\textbf {\bibinfo {volume} {13}},\ \bibinfo {pages} {21} (\bibinfo {year} {1988})}\BibitemShut {NoStop}%
\bibitem [{\citenamefont {Zhu}(1993)}]{zhu1993steady}%
  \BibitemOpen
  \bibfield  {author} {\bibinfo {author} {\bibfnamefont {S.}~\bibnamefont {Zhu}},\ }\bibfield  {title} {\bibinfo {title} {{Steady-State Analysis of a Single-Mode Laser With Correlations Between Additive and Multiplicative Noise}},\ }\href@noop {} {\bibfield  {journal} {\bibinfo  {journal} {Phys. Rev. A}\ }\textbf {\bibinfo {volume} {47}},\ \bibinfo {pages} {2405} (\bibinfo {year} {1993})}\BibitemShut {NoStop}%
\bibitem [{\citenamefont {Chowdhury}\ \emph {et~al.}(2022)\citenamefont {Chowdhury}, \citenamefont {Ray}, \citenamefont {Dana},\ and\ \citenamefont {Ghosh}}]{chowdhury2022extreme}%
  \BibitemOpen
  \bibfield  {author} {\bibinfo {author} {\bibfnamefont {S.~N.}\ \bibnamefont {Chowdhury}}, \bibinfo {author} {\bibfnamefont {A.}~\bibnamefont {Ray}}, \bibinfo {author} {\bibfnamefont {S.~K.}\ \bibnamefont {Dana}},\ and\ \bibinfo {author} {\bibfnamefont {D.}~\bibnamefont {Ghosh}},\ }\bibfield  {title} {\bibinfo {title} {{Extreme Events in Dynamical Systems and Random Walkers: A Review}},\ }\href@noop {} {\bibfield  {journal} {\bibinfo  {journal} {Physics Reports}\ }\textbf {\bibinfo {volume} {966}},\ \bibinfo {pages} {1} (\bibinfo {year} {2022})}\BibitemShut {NoStop}%
\bibitem [{\citenamefont {Schertzer}\ and\ \citenamefont {Lovejoy}(1987)}]{schertzer1987physical}%
  \BibitemOpen
  \bibfield  {author} {\bibinfo {author} {\bibfnamefont {D.}~\bibnamefont {Schertzer}}\ and\ \bibinfo {author} {\bibfnamefont {S.}~\bibnamefont {Lovejoy}},\ }\bibfield  {title} {\bibinfo {title} {{Physical Modeling and Analysis of Rain and Clouds by Anisotropic Scaling Multiplicative Processes}},\ }\href@noop {} {\bibfield  {journal} {\bibinfo  {journal} {J. Geophys. Res.: Atmos.}\ }\textbf {\bibinfo {volume} {92}},\ \bibinfo {pages} {9693} (\bibinfo {year} {1987})}\BibitemShut {NoStop}%
\bibitem [{\citenamefont {Lovejoy}(2018)}]{lovejoy2018spectra}%
  \BibitemOpen
  \bibfield  {author} {\bibinfo {author} {\bibfnamefont {S.}~\bibnamefont {Lovejoy}},\ }\bibfield  {title} {\bibinfo {title} {{Spectra, Intermittency, and Extremes of Weather, Macroweather and Climate}},\ }\href@noop {} {\bibfield  {journal} {\bibinfo  {journal} {Scientific reports}\ }\textbf {\bibinfo {volume} {8}},\ \bibinfo {pages} {12697} (\bibinfo {year} {2018})}\BibitemShut {NoStop}%
\bibitem [{\citenamefont {Ashwin}\ \emph {et~al.}(2012)\citenamefont {Ashwin}, \citenamefont {Wieczorek}, \citenamefont {Vitolo},\ and\ \citenamefont {Cox}}]{ashwin2012tipping}%
  \BibitemOpen
  \bibfield  {author} {\bibinfo {author} {\bibfnamefont {P.}~\bibnamefont {Ashwin}}, \bibinfo {author} {\bibfnamefont {S.}~\bibnamefont {Wieczorek}}, \bibinfo {author} {\bibfnamefont {R.}~\bibnamefont {Vitolo}},\ and\ \bibinfo {author} {\bibfnamefont {P.}~\bibnamefont {Cox}},\ }\bibfield  {title} {\bibinfo {title} {{Tipping Points in Open Systems: Bifurcation, Noise-Induced and Rate-Dependent Examples in the Climate System}},\ }\href@noop {} {\bibfield  {journal} {\bibinfo  {journal} {Phil. Trans. R. Soc. A}\ }\textbf {\bibinfo {volume} {370}},\ \bibinfo {pages} {1166} (\bibinfo {year} {2012})}\BibitemShut {NoStop}%
\bibitem [{\citenamefont {Sura}(2002)}]{sura2002noise}%
  \BibitemOpen
  \bibfield  {author} {\bibinfo {author} {\bibfnamefont {P.}~\bibnamefont {Sura}},\ }\bibfield  {title} {\bibinfo {title} {{Noise-Induced Transitions in a Barotropic $\beta$-Plane Channel}},\ }\href@noop {} {\bibfield  {journal} {\bibinfo  {journal} {J. Atmos. Sci.}\ }\textbf {\bibinfo {volume} {59}},\ \bibinfo {pages} {97} (\bibinfo {year} {2002})}\BibitemShut {NoStop}%
\bibitem [{\citenamefont {Phillips}(2023)}]{syncphillips23}%
  \BibitemOpen
  \bibfield  {author} {\bibinfo {author} {\bibfnamefont {E.~T.}\ \bibnamefont {Phillips}},\ }\bibfield  {title} {\bibinfo {title} {{{The Synchronizing Role of Multiplexing Noise: Exploring Kuramoto Oscillators and Breathing Chimeras}}},\ }\href@noop {} {\bibfield  {journal} {\bibinfo  {journal} {Chaos}\ }\textbf {\bibinfo {volume} {33}},\ \bibinfo {pages} {073140} (\bibinfo {year} {2023})}\BibitemShut {NoStop}%
\bibitem [{\citenamefont {Gilpin}(2021)}]{gilpin2021desynchronization}%
  \BibitemOpen
  \bibfield  {author} {\bibinfo {author} {\bibfnamefont {W.}~\bibnamefont {Gilpin}},\ }\bibfield  {title} {\bibinfo {title} {{Desynchronization of Jammed Oscillators by Avalanches}},\ }\href@noop {} {\bibfield  {journal} {\bibinfo  {journal} {Phys. Rev. Res.}\ }\textbf {\bibinfo {volume} {3}},\ \bibinfo {pages} {023206} (\bibinfo {year} {2021})}\BibitemShut {NoStop}%
\bibitem [{\citenamefont {Schenzle}\ and\ \citenamefont {Brand}(1979)}]{schenzle1979multiplicative}%
  \BibitemOpen
  \bibfield  {author} {\bibinfo {author} {\bibfnamefont {A.}~\bibnamefont {Schenzle}}\ and\ \bibinfo {author} {\bibfnamefont {H.}~\bibnamefont {Brand}},\ }\bibfield  {title} {\bibinfo {title} {{Multiplicative Stochastic Processes in Statistical Physics}},\ }\href@noop {} {\bibfield  {journal} {\bibinfo  {journal} {Phys. Rev. A}\ }\textbf {\bibinfo {volume} {20}},\ \bibinfo {pages} {1628} (\bibinfo {year} {1979})}\BibitemShut {NoStop}%
\bibitem [{\citenamefont {Auma{\^\i}tre}\ \emph {et~al.}(2007)\citenamefont {Auma{\^\i}tre}, \citenamefont {Mallick},\ and\ \citenamefont {P{\'e}tr{\'e}lis}}]{aumaitre2007noise}%
  \BibitemOpen
  \bibfield  {author} {\bibinfo {author} {\bibfnamefont {S.}~\bibnamefont {Auma{\^\i}tre}}, \bibinfo {author} {\bibfnamefont {K.}~\bibnamefont {Mallick}},\ and\ \bibinfo {author} {\bibfnamefont {F.}~\bibnamefont {P{\'e}tr{\'e}lis}},\ }\bibfield  {title} {\bibinfo {title} {{Noise-Induced Bifurcations, Multiscaling and On–Off Intermittency}},\ }\href@noop {} {\bibfield  {journal} {\bibinfo  {journal} {J. Stat. Mech: Theory Exp.}\ }\textbf {\bibinfo {volume} {2007}},\ \bibinfo {pages} {P07016} (\bibinfo {year} {2007})}\BibitemShut {NoStop}%
\bibitem [{\citenamefont {Bourret}\ \emph {et~al.}(1973)\citenamefont {Bourret}, \citenamefont {Frisch},\ and\ \citenamefont {Pouquet}}]{bourret1973brownian}%
  \BibitemOpen
  \bibfield  {author} {\bibinfo {author} {\bibfnamefont {R.}~\bibnamefont {Bourret}}, \bibinfo {author} {\bibfnamefont {U.}~\bibnamefont {Frisch}},\ and\ \bibinfo {author} {\bibfnamefont {A.}~\bibnamefont {Pouquet}},\ }\bibfield  {title} {\bibinfo {title} {{Brownian Motion of Harmonic Oscillator With Stochastic Frequency}},\ }\href@noop {} {\bibfield  {journal} {\bibinfo  {journal} {Physica}\ }\textbf {\bibinfo {volume} {65}},\ \bibinfo {pages} {303} (\bibinfo {year} {1973})}\BibitemShut {NoStop}%
\bibitem [{\citenamefont {Lindenberg}\ \emph {et~al.}(1981)\citenamefont {Lindenberg}, \citenamefont {Seshadri},\ and\ \citenamefont {West}}]{lindenberg1981brownian}%
  \BibitemOpen
  \bibfield  {author} {\bibinfo {author} {\bibfnamefont {K.}~\bibnamefont {Lindenberg}}, \bibinfo {author} {\bibfnamefont {V.}~\bibnamefont {Seshadri}},\ and\ \bibinfo {author} {\bibfnamefont {B.~J.}\ \bibnamefont {West}},\ }\bibfield  {title} {\bibinfo {title} {{Brownian Motion of Harmonic Systems With Fluctuating Parameters: III Scaling and Moment Instabilities}},\ }\href@noop {} {\bibfield  {journal} {\bibinfo  {journal} {Physica A}\ }\textbf {\bibinfo {volume} {105}},\ \bibinfo {pages} {445} (\bibinfo {year} {1981})}\BibitemShut {NoStop}%
\bibitem [{\citenamefont {L{\"u}cke}\ and\ \citenamefont {Schank}(1985)}]{lucke1985response}%
  \BibitemOpen
  \bibfield  {author} {\bibinfo {author} {\bibfnamefont {M.}~\bibnamefont {L{\"u}cke}}\ and\ \bibinfo {author} {\bibfnamefont {F.}~\bibnamefont {Schank}},\ }\bibfield  {title} {\bibinfo {title} {{Response to Parametric Modulation Near an Instability}},\ }\href@noop {} {\bibfield  {journal} {\bibinfo  {journal} {Phys. Rev. Lett.}\ }\textbf {\bibinfo {volume} {54}},\ \bibinfo {pages} {1465} (\bibinfo {year} {1985})}\BibitemShut {NoStop}%
\bibitem [{\citenamefont {Graham}\ and\ \citenamefont {Schenzle}(1982)}]{graham1982stabilization}%
  \BibitemOpen
  \bibfield  {author} {\bibinfo {author} {\bibfnamefont {R.}~\bibnamefont {Graham}}\ and\ \bibinfo {author} {\bibfnamefont {A.}~\bibnamefont {Schenzle}},\ }\bibfield  {title} {\bibinfo {title} {{Stabilization by Multiplicative Noise}},\ }\href@noop {} {\bibfield  {journal} {\bibinfo  {journal} {Phys. Rev. A}\ }\textbf {\bibinfo {volume} {26}},\ \bibinfo {pages} {1676} (\bibinfo {year} {1982})}\BibitemShut {NoStop}%
\bibitem [{\citenamefont {Arnold}\ \emph {et~al.}(1983)\citenamefont {Arnold}, \citenamefont {Crauel},\ and\ \citenamefont {Wihstutz}}]{arnold1983stabilization}%
  \BibitemOpen
  \bibfield  {author} {\bibinfo {author} {\bibfnamefont {L.}~\bibnamefont {Arnold}}, \bibinfo {author} {\bibfnamefont {H.}~\bibnamefont {Crauel}},\ and\ \bibinfo {author} {\bibfnamefont {V.}~\bibnamefont {Wihstutz}},\ }\bibfield  {title} {\bibinfo {title} {{Stabilization of Linear Systems by Noise}},\ }\href@noop {} {\bibfield  {journal} {\bibinfo  {journal} {SIAM J. Control Optim.}\ }\textbf {\bibinfo {volume} {21}},\ \bibinfo {pages} {451} (\bibinfo {year} {1983})}\BibitemShut {NoStop}%
\bibitem [{\citenamefont {Arnold}\ \emph {et~al.}(1979)\citenamefont {Arnold}, \citenamefont {Horsthemke},\ and\ \citenamefont {Stucki}}]{arnold1979influence}%
  \BibitemOpen
  \bibfield  {author} {\bibinfo {author} {\bibfnamefont {L.}~\bibnamefont {Arnold}}, \bibinfo {author} {\bibfnamefont {W.}~\bibnamefont {Horsthemke}},\ and\ \bibinfo {author} {\bibfnamefont {J.}~\bibnamefont {Stucki}},\ }\bibfield  {title} {\bibinfo {title} {{The Influence of External Real and White Noise on the Lotka-Volterra Model}},\ }\href@noop {} {\bibfield  {journal} {\bibinfo  {journal} {Biom. J.}\ }\textbf {\bibinfo {volume} {21}},\ \bibinfo {pages} {451} (\bibinfo {year} {1979})}\BibitemShut {NoStop}%
\bibitem [{\citenamefont {Bobrik}\ and\ \citenamefont {Stettner}(1999)}]{bobrik1999stabilizing}%
  \BibitemOpen
  \bibfield  {author} {\bibinfo {author} {\bibfnamefont {R.}~\bibnamefont {Bobrik}}\ and\ \bibinfo {author} {\bibfnamefont {L.}~\bibnamefont {Stettner}},\ }\bibfield  {title} {\bibinfo {title} {{The Stabilizing Effect of a Random Parametric Perturbation on an Unstable Linear System}},\ }\href@noop {} {\bibfield  {journal} {\bibinfo  {journal} {J. Math. Sci.}\ }\textbf {\bibinfo {volume} {96}},\ \bibinfo {pages} {3038} (\bibinfo {year} {1999})}\BibitemShut {NoStop}%
\bibitem [{\citenamefont {Kaulakys}\ \emph {et~al.}(2006)\citenamefont {Kaulakys}, \citenamefont {Ruseckas}, \citenamefont {Gontis},\ and\ \citenamefont {Alaburda}}]{kaulakys2006nonlinear}%
  \BibitemOpen
  \bibfield  {author} {\bibinfo {author} {\bibfnamefont {B.}~\bibnamefont {Kaulakys}}, \bibinfo {author} {\bibfnamefont {J.}~\bibnamefont {Ruseckas}}, \bibinfo {author} {\bibfnamefont {V.}~\bibnamefont {Gontis}},\ and\ \bibinfo {author} {\bibfnamefont {M.}~\bibnamefont {Alaburda}},\ }\bibfield  {title} {\bibinfo {title} {{Nonlinear Stochastic Models of 1/F Noise and Power-Law Distributions}},\ }\href@noop {} {\bibfield  {journal} {\bibinfo  {journal} {Phys. A}\ }\textbf {\bibinfo {volume} {365}},\ \bibinfo {pages} {217} (\bibinfo {year} {2006})}\BibitemShut {NoStop}%
\bibitem [{\citenamefont {Kaulakys}\ and\ \citenamefont {Alaburda}(2009)}]{kaulakys2009modeling}%
  \BibitemOpen
  \bibfield  {author} {\bibinfo {author} {\bibfnamefont {B.}~\bibnamefont {Kaulakys}}\ and\ \bibinfo {author} {\bibfnamefont {M.}~\bibnamefont {Alaburda}},\ }\bibfield  {title} {\bibinfo {title} {{Modeling Scaled Processes and 1/F$\beta$ Noise Using Nonlinear Stochastic Differential Equations}},\ }\href@noop {} {\bibfield  {journal} {\bibinfo  {journal} {J. Stat. Mech: Theory Exp.}\ }\textbf {\bibinfo {volume} {2009}},\ \bibinfo {pages} {P02051} (\bibinfo {year} {2009})}\BibitemShut {NoStop}%
\bibitem [{\citenamefont {Hramov}\ \emph {et~al.}(2007)\citenamefont {Hramov}, \citenamefont {Koronovskii}, \citenamefont {Kurovskaya}, \citenamefont {Ovchinnikov},\ and\ \citenamefont {Boccaletti}}]{hramov2007length}%
  \BibitemOpen
  \bibfield  {author} {\bibinfo {author} {\bibfnamefont {A.~E.}\ \bibnamefont {Hramov}}, \bibinfo {author} {\bibfnamefont {A.~A.}\ \bibnamefont {Koronovskii}}, \bibinfo {author} {\bibfnamefont {M.~K.}\ \bibnamefont {Kurovskaya}}, \bibinfo {author} {\bibfnamefont {A.~A.}\ \bibnamefont {Ovchinnikov}},\ and\ \bibinfo {author} {\bibfnamefont {S.}~\bibnamefont {Boccaletti}},\ }\bibfield  {title} {\bibinfo {title} {{Length Distribution of Laminar Phases for Type-I Intermittency in the Presence of Noise}},\ }\href@noop {} {\bibfield  {journal} {\bibinfo  {journal} {Phys. Rev. E}\ }\textbf {\bibinfo {volume} {76}},\ \bibinfo {pages} {026206} (\bibinfo {year} {2007})}\BibitemShut {NoStop}%
\bibitem [{\citenamefont {Hirsch}\ \emph {et~al.}(1982)\citenamefont {Hirsch}, \citenamefont {Huberman},\ and\ \citenamefont {Scalapino}}]{hirsch1982theory}%
  \BibitemOpen
  \bibfield  {author} {\bibinfo {author} {\bibfnamefont {J.}~\bibnamefont {Hirsch}}, \bibinfo {author} {\bibfnamefont {B.}~\bibnamefont {Huberman}},\ and\ \bibinfo {author} {\bibfnamefont {D.}~\bibnamefont {Scalapino}},\ }\bibfield  {title} {\bibinfo {title} {{Theory of Intermittency}},\ }\href@noop {} {\bibfield  {journal} {\bibinfo  {journal} {Phys. Rev. A}\ }\textbf {\bibinfo {volume} {25}},\ \bibinfo {pages} {519} (\bibinfo {year} {1982})}\BibitemShut {NoStop}%
\bibitem [{\citenamefont {Kye}\ and\ \citenamefont {Kim}(2000)}]{kye2000characteristic}%
  \BibitemOpen
  \bibfield  {author} {\bibinfo {author} {\bibfnamefont {W.-H.}\ \bibnamefont {Kye}}\ and\ \bibinfo {author} {\bibfnamefont {C.-M.}\ \bibnamefont {Kim}},\ }\bibfield  {title} {\bibinfo {title} {{Characteristic Relations of Type-I Intermittency in the Presence of Noise}},\ }\href@noop {} {\bibfield  {journal} {\bibinfo  {journal} {Phys. Rev. E}\ }\textbf {\bibinfo {volume} {62}},\ \bibinfo {pages} {6304} (\bibinfo {year} {2000})}\BibitemShut {NoStop}%
\bibitem [{\citenamefont {Gardiner}(2009)}]{gardiner2009stochastic}%
  \BibitemOpen
  \bibfield  {author} {\bibinfo {author} {\bibfnamefont {C.}~\bibnamefont {Gardiner}},\ }\href@noop {} {\emph {\bibinfo {title} {{Stochastic Methods}}}},\ Vol.~\bibinfo {volume} {4}\ (\bibinfo  {publisher} {Springer Berlin},\ \bibinfo {year} {2009})\BibitemShut {NoStop}%
\bibitem [{\citenamefont {Ahlers}\ \emph {et~al.}(1984)\citenamefont {Ahlers}, \citenamefont {Hohenberg},\ and\ \citenamefont {L{\"u}cke}}]{ahlers1984externally}%
  \BibitemOpen
  \bibfield  {author} {\bibinfo {author} {\bibfnamefont {G.}~\bibnamefont {Ahlers}}, \bibinfo {author} {\bibfnamefont {P.}~\bibnamefont {Hohenberg}},\ and\ \bibinfo {author} {\bibfnamefont {M.}~\bibnamefont {L{\"u}cke}},\ }\bibfield  {title} {\bibinfo {title} {{Externally Modulated Rayleigh-B{\'e}nard Convection: Experiment and Theory}},\ }\href@noop {} {\bibfield  {journal} {\bibinfo  {journal} {Phys. Rev. Lett.}\ }\textbf {\bibinfo {volume} {53}},\ \bibinfo {pages} {48} (\bibinfo {year} {1984})}\BibitemShut {NoStop}%
\bibitem [{\citenamefont {Lindner}\ \emph {et~al.}(2001)\citenamefont {Lindner}, \citenamefont {Kostur},\ and\ \citenamefont {Schimansky-Geier}}]{lindner2001optimal}%
  \BibitemOpen
  \bibfield  {author} {\bibinfo {author} {\bibfnamefont {B.}~\bibnamefont {Lindner}}, \bibinfo {author} {\bibfnamefont {M.}~\bibnamefont {Kostur}},\ and\ \bibinfo {author} {\bibfnamefont {L.}~\bibnamefont {Schimansky-Geier}},\ }\bibfield  {title} {\bibinfo {title} {{Optimal Diffusive Transport in a Tilted Periodic Potential}},\ }\href@noop {} {\bibfield  {journal} {\bibinfo  {journal} {Fluctuation Noise Lett.}\ }\textbf {\bibinfo {volume} {1}},\ \bibinfo {pages} {R25} (\bibinfo {year} {2001})}\BibitemShut {NoStop}%
\bibitem [{\citenamefont {Levy}\ and\ \citenamefont {Solomon}(1996)}]{levy1996power}%
  \BibitemOpen
  \bibfield  {author} {\bibinfo {author} {\bibfnamefont {M.}~\bibnamefont {Levy}}\ and\ \bibinfo {author} {\bibfnamefont {S.}~\bibnamefont {Solomon}},\ }\bibfield  {title} {\bibinfo {title} {{Power Laws Are Logarithmic Boltzmann Laws}},\ }\href@noop {} {\bibfield  {journal} {\bibinfo  {journal} {Int. J. Mod. Phys. C}\ }\textbf {\bibinfo {volume} {7}},\ \bibinfo {pages} {595} (\bibinfo {year} {1996})}\BibitemShut {NoStop}%
\bibitem [{\citenamefont {Sornette}\ and\ \citenamefont {Cont}(1997)}]{sornette1997convergent}%
  \BibitemOpen
  \bibfield  {author} {\bibinfo {author} {\bibfnamefont {D.}~\bibnamefont {Sornette}}\ and\ \bibinfo {author} {\bibfnamefont {R.}~\bibnamefont {Cont}},\ }\bibfield  {title} {\bibinfo {title} {{Convergent Multiplicative Processes Repelled From Zero: Power Laws and Truncated Power Laws}},\ }\href@noop {} {\bibfield  {journal} {\bibinfo  {journal} {J. Phys. I}\ }\textbf {\bibinfo {volume} {7}},\ \bibinfo {pages} {431} (\bibinfo {year} {1997})}\BibitemShut {NoStop}%
\bibitem [{\citenamefont {Sornette}(1998)}]{sornette1998multiplicative}%
  \BibitemOpen
  \bibfield  {author} {\bibinfo {author} {\bibfnamefont {D.}~\bibnamefont {Sornette}},\ }\bibfield  {title} {\bibinfo {title} {{Multiplicative Processes and Power Laws}},\ }\href@noop {} {\bibfield  {journal} {\bibinfo  {journal} {Phys. Rev. E}\ }\textbf {\bibinfo {volume} {57}},\ \bibinfo {pages} {4811} (\bibinfo {year} {1998})}\BibitemShut {NoStop}%
\end{thebibliography}%

\end{document}